\documentclass[12pt]{article}
\usepackage{graphicx}
\usepackage{amssymb}
\usepackage{amsmath}
\newcommand{\vev}[1]{{\langle #1 \rangle}}
\newcommand{\abs}[1]{{\left| #1 \right|}}

\newcommand{\eV}{\mbox{~eV}}
\newcommand{\MeV}{\mbox{~MeV}}
\newcommand{\GeV}{\mbox{~GeV}}

\newcommand{\ie}{{\it i.e.}}
\newcommand{\eg}{{\it e.g.}}
\newcommand{\eqn}[1]{&\hspace{-0.6em}#1\hspace{-0.6em}&}
\begin{document}
\baselineskip 0.6cm
%
\begin{titlepage}
\begin{center}

\begin{flushright}
\end{flushright}

\vskip 2cm

{\Large \bf Mixing of Active and Sterile Neutrinos}

\vskip 1.2 cm

{\large 
Takehiko Asaka$^{1,2}$, Shintaro Eijima$^3$ and Hiroyuki Ishida$^3$
}

\vskip 0.4cm

$^1${\em
  Department of Physics, Niigata University, Niigata 950-2181, Japan
}

$^2${\em
  Max-Planck-Institut f\"ur Kernphysik,
  Postfach 103980, 69029 Heidelberg, Germany
}

$^3${\em
  Graduate School of Science and Technology, Niigata University, Niigata 950-2181, Japan
}

\vskip 0.2cm

\today

\vskip 2cm

\vskip .5in

\begin{abstract}
  We investigate mixing of neutrinos in the $\nu$MSM (neutrino Minimal
  Standard Model), which is the MSM extended by three right-handed
  neutrinos.  Especially, we study elements of the mixing matrix
  $\Theta_{\alpha I}$ between three left-handed neutrinos $\nu_\alpha$
  ($\alpha = e,\mu,\tau$) and two sterile neutrinos $N_I$ ($I=2,3$)
  which are responsible to the seesaw mechanism generating the
  suppressed masses of active neutrinos as well as the generation of
  the baryon asymmetry of the universe (BAU).  It is shown that
  $\Theta_{eI}$ can be suppressed by many orders of magnitude compared
  with $\Theta_{\mu I}$ and $\Theta_{\tau I}$, when the Chooz angle
  $\theta_{13}$ is large in the normal hierarchy of active neutrino
  masses.  We then discuss the neutrinoless double beta decay in this
  framework by taking into account the contributions not only from
  active neutrinos but also from all the three sterile neutrinos.  It
  is shown that $N_2$ and $N_3$ give substantial, destructive contributions when
  their masses are smaller than a few 100 MeV, and as a results
  $\Theta_{e I}$ receive no stringent constraint from the current
  bounds on such decay.  Finally, we discuss the impacts of the
  obtained results on the direct searches of $N_{2,3}$ in meson decays
  for the case when $N_{2,3}$ are lighter than pion mass.  We show
  that there exists the allowed region for $N_{2,3}$ with such small
  masses in the normal hierarchy case even if the current bound on the
  lifetimes of $N_{2,3}$ from the big bang nucleosynthesis is imposed.
  It is also pointed out that the direct search by using $\pi^+ \to
  e^+ + N_{2,3}$ and $K^+ \to e^+ + N_{2,3}$ might miss such $N_{2,3}$
  since the branching ratios can be extremely small due to the
  cancellation in $\Theta_{eI}$, but the search by $K^+ \to \mu^+ +
  N_{2,3}$ can cover the whole allowed region by improving the
  measurement of the branching ratio by a factor of 5.
\end{abstract}
\end{center}
\end{titlepage}
\renewcommand{\thefootnote}{\#\arabic{footnote}} 
%
\section{Introduction}
\label{sec:introduction}
The extension by right-handed neutrinos is one of the most interesting
physics beyond the Minimal Standard Model (MSM), since it gives a
simple solution to the problem of the neutrino masses confirmed by
various oscillation experiments.  Usually, right-handed neutrinos are
introduced with superheavy Majorana masses and sizable Yukawa coupling
constants in order to realize the seesaw mechanism~\cite{Seesaw},
which accounts naturally for the smallness of neutrino masses.
Furthermore, the decays of such right-handed neutrinos can be a source
of the baryon asymmetry of the universe (BAU) through the
leptogenesis mechanism~\cite{Fukugita:1986hr,Leptogenesis}.  When the
masses of right-handed neutrinos are hierarchical, the observed BAU
requires the mass of the lightest one should be heavier than about
$10^9$ GeV~\cite{Giudice:2003jh}.  Although such singlet fermions
provide simple and natural solution to the origins of neutrino masses
and BAU at the same time, it is almost impossible to test them at
experiments  in the future.

It should, however, be noted that right-handed neutrinos can bring
about important physical phenomena, even when the scale of Majorana
masses are so light to be produced in terrestrial experiments.  One
attractive possibility is the
$\nu$MSM~\cite{Asaka:2005an,Asaka:2005pn} in which three right-handed
neutrinos are introduced with masses below the electroweak scale.%
\footnote{
The explanation of the LSND anomaly in this framework 
had been investigated in Ref.~\cite{deGouvea:2005er}.
The non-minimal coupling of the Higgs field to gravity
allows to realize the cosmic inflation~\cite{Bezrukov:2007ep}.
Further, various extensions of the model have been 
discussed~\cite{Shaposhnikov:2006xi}-\cite{Bezrukov:2008ut}.}
Interestingly, this simple model can explain the origins of neutrino
masses, BAU and also dark matter of the universe at the same time.
The Yukawa coupling constants are so small that the seesaw mechanism
still works, and mass eigenstates of neutrinos are divided into two
groups, active and sterile neutrinos.  The flavour mixing of active
neutrinos accounts for neutrino oscillations observed in experiments.
On the other hand, three sterile neutrinos, $N_1$, $N_2$, and $N_3$,
solve cosmological problems in the MSM.

One of the sterile neutrinos, $N_1$, plays a role of dark
matter~\cite{SterileNeutrinoDM}.%
\footnote{%
  It has been investigated various phenomenon in astrophysics by the
  sterile neutrino dark-matter, \eg, the explanation of the pulsar
  kick~\cite{Pulsarkick}.  See, for example, a review of this
  issue~\cite{Kusenko:2009up} and references therein.}  
When its mass is in the keV order, it can be produced by the so-called
Dodelson-Widrow mechanism~\cite{Dodelson:1993je}, \ie, by the thermal
scatterings through its mixing with the left-handed neutrinos.  See
Refs.~\cite{Shi:1998km}--\cite{Laine:2008pg}.
This dark-matter particle receives severe astrophysical
constraints~\cite{Boyarsky:2009ix}.  One important bound comes from
the $X$-ray 
background~\cite{Dolgov:2000ew}--\cite{Abazajian:2001vt} 
and the other comes from the cosmic structure at small scales like the
Lyman $\alpha$ 
forest~\cite{Viel:2006kd}--\cite{Boyarsky:2008mt}.  Even when these
constraints are imposed, as shown in Ref.~\cite{Laine:2008pg}, the
correct abundance of the dark matter can be obtained through the
Dodelson-Widrow mechanism by invoking the resonant
production~\cite{Shi:1998km} in the presence of the large lepton
asymmetry.  Notice that $N_1$ plays no significant role in the seesaw
mechanism~\cite{Asaka:2005an}.  This is because Yukawa coupling
constants of $N_1$ is so small that its contribution to the mass
matrix of active neutrinos is negligible.  Due to the very suppressed
interaction the direct search of $N_1$ at experiments is very
difficult.  However, it can be observed by the specific spectrum in
the X-ray background coming from the decay of $N_1$ into active
neutrino and photon.  It has also been discussed the search in
laboratory~\cite{Bezrukov:2006cy}.

The rest two sterile neutrinos, $N_2$ and $N_3$, are responsible to
generate not only the seesaw masses of active neutrinos but also BAU
through the mechanism~\cite{Akhmedov:1998qx}.  The flavour oscillation
between $N_2$ and $N_3$ in the early universe induces the separation
of lepton asymmetry between left- and right-handed leptons and the
asymmetry in the left-handed sector is partially converted into the
baryon asymmetry by the sphaleron processes~\cite{Kuzmin:1985mm}.  The
$\nu$MSM realizes this baryogenesis scenario without conflict with
the observational data of the neutrino oscillations when their masses
are quasi-degenerate and in the range ${\cal O}(0.1)$--${\cal O}(10)$
GeV~\cite{Asaka:2005pn}.  See also the recent analysis in
Refs.~\cite{Asaka:2010kk,Canetti:2010aw}.  

It is interesting to note that sterile neutrinos $N_2$ and $N_3$ can
be tested in various experiments as pointed out in
Ref.~\cite{Gorbunov:2007ak}.  This is crucially important to reveal
the origins of the neutrino masses as well as the cosmic baryon
asymmetry.  For this purpose, we would like to study the mixing of
sterile neutrinos with left-handed neutrinos $\nu_{\alpha}$ ($\alpha =
e, \mu, \tau$) in this paper.  The elements of such mixing matrix,
$\Theta_{\alpha I},$ are vital to discuss phenomenology of the
$\nu$MSM, since the strength of the interactions of sterile neutrinos
is determined by them.  Sterile neutrinos in the model possess the
Yukawa interactions and also the weak gauge interactions via the above
mixing after the electroweak symmetry breaking. Since the elements
$\Theta_{\alpha I}$ are proportional to the Yukawa coupling constants
$F_{\alpha I}$, both interactions are controlled by $\Theta_{\alpha
  I}$.

It should be noted that the mixing elements of $N_2$ and $N_3$ can
take values  varying by many orders of magnitude.
This point had already been discussed by using the
model with the lepton symmetry~\cite{Shaposhnikov:2006nn}.  As we will
show in Sec.~\ref{sec:mixing}, the mixing elements increase
exponentially as $\Theta_{\alpha I} \propto \exp( \mbox{Im}\omega )$
for large $\mbox{Im} \omega$ ($\omega$ is a complex parameter in the
neutrino Yukawa matrix) keeping the masses and mixing angles of active
neutrinos unchanged.  This enhancement leads to the various 
significant impacts on phenomenology of sterile neutrinos.
For example,  it gives the larger production/detection rates in the
search experiments, the larger CP asymmetry in baryogenesis,
and the shorter lifetime of $N_{2,3}$ making them cosmologically 
harmless.

The purpose of this paper is, thus, to investigate the mixing elements
$\Theta_{\alpha I}$ in detail, and to reveal how the elements of
$N_{2}$ and $N_{3}$ depend on mass hierarchy, mixing angles and CP
phases of active neutrinos in addition to the parameters of sterile
neutrinos.  Our analysis will show that there can be strong hierarchy
among the mixing elements $\Theta_{eI}$, $\Theta_{\mu I}$ and
$\Theta_{\tau I}$ depending on choice of the parameters.  Especially,
it will been pointed out that the strong suppression in the mixing
elements of electron type $\Theta_{e I}$ can happen for the normal
hierarchy of active neutrino masses, which enlarges the allowed region
of the model.  Although we shall consider the $\nu$MSM, the results of
the mixing elements $\Theta_{\alpha I}$ in this paper can be applied
to the general models of the seesaw mechanism with two right-handed
neutrinos.

Further, we would like to discuss the implications to two phenomena of
sterile neutrinos.  The first one is the neutrinoless double beta
decay in which the mixing elements of active and sterile neutrinos
play the crucial roles.  This problem had already been discussed in
Ref.~\cite{Bezrukov:2005mx}.  We will extend the analysis especially
when the masses of $N_2$ and $N_3$ are smaller than a few 100 MeV, and
show that the rates of the decays in the $\nu$MSM is smaller than
those in the usual case where only active neutrinos give the
contribution.  Thus, the model is free from significant constraints
discussed in Ref.~\cite{Benes:2005hn}.  
The other one is the search of
$N_{2,3}$ produced in the decays of charged pions and/or kaons where
the elements $|\Theta_{e I}|^2$ or $|\Theta_{\mu I}|^2$ determine the
production rates.   Especially, we will point out that
$N_2$ and $N_3$ which masses are smaller than pion mass
are still allowed by the constraints from the direct searches
as well as that from the big bang 
nucleosynthesis~\cite{Dolgov:2000pj,Dolgov:2000jw}
for the normal hierarchy case.
This is different from a conclusion from Ref.~\cite{Gorbunov:2007ak}.
The reason for it lies in the cancellation in the $\Theta_{eI}$
mentioned above.  In addition, we shall present some implications
to the future searches by using the decays of $\pi^+$ and $K^+$.

This paper is organized as follows.  In Sec.~\ref{sec:model} we
briefly review the framework of the present analysis, \ie, the
$\nu$MSM.  In Sec.~\ref{sec:mixing} we study the mixing of sterile
neutrinos $N_2$ and $N_3$ in the charged current interactions.
Especially, we investigate how the mixing elements depend on the
parameters of active neutrinos, \ie, the mass hierarchy, mixing angles
and CP violating phases.  We then apply the obtained results in
phenomenology of $N_2$ and $N_3$.  In Sec.~\ref{sec:0nu2beta} we
estimate the contributions of sterile neutrinos to the neutrinoless
double beta decay and address the importance of such contributions
when the masses of $N_2$ and $N_3$ are smaller than about the order of
100 MeV.  In Sec.~\ref{sec:search} we discuss search of $N_2$ and
$N_3$ in the charged pion and kaon decays for the case when their
masses are lighter than the pion mass.  Finally, our results are
summarized in Sec.~\ref{sec:conc}.  We add App.~\ref{sec:appendix} to
present the expressions for the mixing elements.

\section{The $\nu$MSM}
\label{sec:model}
First of all, we review the $\nu$MSM, which is the MSM extended by
three right-handed neutrinos $\nu_{R \, I}$ ($I=1,2,3$) 
with Lagrangian
\begin{eqnarray}
  \label{eq:L_nuMSM}
  {\cal L}_{\nu{\rm MSM}} =
  {\cal L}_{\rm MSM} +
  i \, \overline{\nu_R{}_I} \, \gamma^\mu \, \partial_\mu \, \nu_R{}_I
  -
  \Bigl(
  F_{\alpha I} \, \overline{L}_\alpha \, \Phi \, \nu_R{}_I
  + \frac{M_I}{2} \, \overline{\nu_R{}_I^c} \, \nu_R{}_I 
  + h.c.
  \Bigr)
\,,
\end{eqnarray}
where ${\cal L}_{\rm MSM}$ is the MSM Lagrangian.
$\Phi$ and
$L_\alpha = (e_{L \, \alpha}, \nu_{L \, \alpha})^T$ ($\alpha = e, \mu,
\tau$) are Higgs and lepton weak-doublets, respectively.  We denote
Yukawa coupling constants of neutrinos by $F_{\alpha I}$.  Here and
hereafter we work in a basis in which the mass matrix of charged
leptons is diagonal.  In this model neutrinos receive the Majorana
masses $[M_M]_{IJ} = M_I \delta_{IJ}$ (which are taken to be real and
positive without loss of generality) and the Dirac masses
$[M_D]_{\alpha I} = F_{\alpha I} \vev{\Phi}$ ($\vev{\Phi}$ is a vacuum
expectation value of the Higgs field).  The distinctive feature of the
model is the region of the parameter space of Eq.~(\ref{eq:L_nuMSM}),
\ie, we restrict ourselves in the region
\begin{eqnarray}
  \label{eq:hierarchy}
  \left|[M_D]_{\alpha I} \right| \ll M_I \lesssim \Lambda_{\rm EW} \, \,.
\end{eqnarray}
Notice that the seesaw mechanism still works even if the Majorana
masses are smaller than or comparable to the weak scale $\Lambda_{\rm
  EW} = {\cal O}(10^2)$ GeV.  This is simply because neutrino Yukawa
coupling constants of interest are extremely small.  (See the
discussion below.)

The mass matrix of neutrinos $\hat M$, which is a $6 \times 6$
symmetric matrix, is given by
\begin{eqnarray}
  \hat M = 
  \left(
    \begin{array}{c c}
      0 & M_D \\ M_D^T & M_M
    \end{array}
  \right) \,.
\end{eqnarray}
We can diagonalize it by using the unitary matrix $\hat U$ as $\hat
U^\dagger \, \hat M \, \hat U^\ast = \hat M^{\rm diag}$.  The seesaw
mechanism shows that $\hat U$ at the leading order takes the form
\begin{eqnarray}
  \label{eq:Uhat}
  \hat U =
  \left(
    \begin{array}{c c}
      U & \Theta \\ - \Theta^\dagger \, U & 1
    \end{array}
  \right) \,.
\end{eqnarray}
Here $U$ is the $3\times 3$ Pontecorvo-Maki-Nakagawa-Sakata
(PMNS) matrix~\cite{PMNS};
\begin{eqnarray}
  U^\dagger M_\nu U^\ast = \mbox{\rm diag}(m_1, m_2, m_3) \,,
\end{eqnarray}
where $M_\nu = - M_D M_M^{-1} M_D^T$ is the seesaw mass matrix.  We
call the eigenstates having masses $m_i$ as active neutrinos $\nu_i$
($i=1,2,3)$.  The rest three mass eigenstates, denoted by $N_I$, are
almost corresponding to right-handed neutrinos $N_I \simeq \nu_{R \,
  I}$ having masses $M_I$.  The neutrino mixing in the charge current
is then induced through
\begin{eqnarray}
  \nu_L{}_\alpha = 
  U_{\alpha i} \, \nu_i + \Theta_{\alpha I} \, N_I^c \,,
\end{eqnarray}
where the $3 \times 3$ mixing matrix $\Theta$ is 
found at the leading order as
\begin{eqnarray}
  \Theta_{\alpha I} = \frac{[M_D]_{\alpha I}}{M_I} \,,
\end{eqnarray}
and hence $\abs{\Theta_{\alpha I}} \ll 1$ due to
Eq.~(\ref{eq:hierarchy}).
We shall call $N_I$ as sterile neutrinos since
they possess very suppressed gauge interactions.
It should be stressed that
sterile neutrinos here are originated from right-handed
neutrinos in the seesaw mechanism. 

In the $\nu$MSM three right-handed neutrinos play important roles in
cosmology.  One of them, say $N_1$, is a candidate for dark matter of
the universe.  This dark-matter particle receives severe astrophysical
constraints as mentioned in Sec.~\ref{sec:introduction}.  Even then,
the correct dark-matter abundance can be obtained through the
mechanism~\cite{Dodelson:1993je} with the resonant
production~\cite{Shi:1998km} in the presence of the large lepton
asymmetry.  The recent study~\cite{Laine:2008pg} shows that the
required mass is $M_1 = 4$--$50$ keV and the Yukawa coupling constants
are typically $|F_{\alpha 1}| =5 \times 10^{-15}$--$4 \times
10^{-13}$.  As a result, $N_1$ gives no significant contribution to
the seesaw mass matrix $M_\nu$~\cite{Asaka:2005an}.

The other right-handed neutrinos, $N_2$ and $N_3$, are then
responsible to the masses and mixing of active neutrinos.
Notice that in this case the mass of the lightest active neutrino
becomes $m_1 < {\cal O}(10^{-6})$ eV.  Further, $N_2$ and $N_3$ can
explain the origin of BAU.  The flavour oscillation between them in the
early universe can generate BAU through the mechanism proposed in
Ref.~\cite{Akhmedov:1998qx}.  In the $\nu$MSM 
the correct amount of BAU can be obtained
when $N_2$ and $N_3$ are quasi-degenerate in
mass~\cite{Asaka:2005pn,Asaka:2010kk,Canetti:2010aw}.

The main purpose of this paper is to study the 
mixing elements $\Theta_{\alpha I}$ ($I=2,3$)
of sterile neutrinos $N_2$ and $N_3$ with flavour 
neutrinos.
To do this, let us express their Yukawa coupling constants
by using mixing angles and masses of active neutrinos
in oscillation experiments.  
As mentioned before, the successful dark matter scenario
requires very small Yukawa couplings of $N_1$ and
its contribution to $M_\nu$ can be neglected.
Thus, we set $F_{\alpha 1} = 0$ here for simplicity.
(See, however, the discussion in Sec.~\ref{sec:0nu2beta}.)
In this case
the neutrino Yukawa matrix $F$ for $N_2$ and $N_3$, which is a $3
\times 2$ matrix, can be expressed without loss of generality
as~\cite{Casas:2001sr,Abada:2006ea}
\begin{eqnarray}
  \label{eq:F}
    F = \frac{i}{\vev{\Phi}} \,
    U \, D_\nu^{1/2} \, \Omega \, D_N^{1/2} \,.
\end{eqnarray}
Here parameters of active neutrinos are their 
masses $D_\nu = \mbox{diag}(m_1, m_2,m_3)$ and 
the mixing matrix
\begin{eqnarray}
  U = 
  \left( 
    \begin{array}{c c c}
      c_{12} c_{13} &
      s_{12} c_{13} &
      s_{13} e^{- i \delta} 
      \\
      - c_{23} s_{12} - s_{23} c_{12} s_{13} e^{i \delta} &
      c_{23} c_{12} - s_{23} s_{12} s_{13} e^{i \delta} &
      s_{23} c_{13} 
      \\
      s_{23} s_{12} - c_{23} c_{12} s_{13} e^{i \delta} &
      - s_{23} c_{12} - c_{23} s_{12} s_{13} e^{i \delta} &
      c_{23} c_{13}
    \end{array}
  \right)  
  \times
  \mbox{diag} 
  ( 1 \,,~ e^{i \eta} \,,~ 1) \,,
\end{eqnarray}
with $s_{ij} = \sin \theta_{ij}$ and $c_{ij} = \cos \theta_{ij}$.  
Note that there is one Majorana phase $\eta$ in 
addition to Dirac phase $\delta$ under the considering 
situation.  Because we have set $F_{\alpha 1} = 0$, masses of active
neutrinos are
\begin{eqnarray}
  &&m_3 = m_{\rm atm} > m_2 = m_{\rm sol} > m_1 =0
  ~~~\mbox{in the NH case} \,,
  \nonumber \\
  &&m_2 = \sqrt{m_{\rm atm}^2 + m_{\rm sol}^2} > 
  m_1 = \sqrt{m_{\rm atm}^2} > m_3 = 0
  ~~~\mbox{in the IH case} \,,
\end{eqnarray}
The observational data of mixing angles are $s_{12}^2 =
0.318^{+0.062}_{-0.048}$, $s_{23}^2 = 0.50^{+0.17}_{-0.14}$, and
$s_{13}^2 \le 0.053$, respectively,
and masses are 
$m_{\rm sol}^2 = \Delta m_{21}^2 = (7.59^{+0.68}_{-0.56}) \times 10^{-5}
\eV^2$ and 
$m_{\rm atm}^2 = |\Delta m_{31}^2| = (2.40^{+0.35}_{-0.33} )\times
10^{-3}\eV^2$ (at the $3 \sigma$ level)~\cite{Schwetz:2008er}.
Hereafter, we shall adopt the central values
unless otherwise stated.

On the other hand, parameters of $N_2$ and $N_3$ are
their masses $D_N = \mbox{diag}(M_2,M_3)$ and the $3 \times 2$
matrix
\begin{eqnarray}
  &&\Omega =
  \left(
    \begin{array}{c c}
      0 & 0 \\
      \cos \omega & - \sin \omega \\
      \xi \sin \omega & \xi \cos \omega
    \end{array}
  \right)
  ~~\mbox{in the NH case} \,,
  \nonumber \\
  &&\Omega =
  \left(
    \begin{array}{c c}
      \cos \omega & - \sin \omega \\
      \xi \sin \omega & \xi \cos \omega \\
      0 & 0 
    \end{array}
  \right)
  ~~\mbox{in the IH case} \,,
\end{eqnarray}
where $\xi = \pm 1$ and $\omega$ is an arbitrary complex number.  
Notice that the change of the sign $\xi$ can be compensated by 
$\omega \to - \omega$ together with the redefinition of 
$N_3$ as $\xi N_3 \to N_3$~\cite{Abada:2006ea}.

\section{Mixing matrix of sterile neutrinos}
\label{sec:mixing}
%
The important parameters for phenomenology of sterile neutrinos $N_2$
and $N_3$ are their masses $M_{2,3}$ and mixing matrix $\Theta$.
Especially, the latter one is crucial to specify the strength of
interactions with other particles.  Here we would like to discuss how
they depends on the parameters of active neutrinos.

Before discussing the $\nu$MSM, let us consider a toy model with one
pair of left- and right-handed neutrinos.  In this case, the mixing of
sterile neutrino is given by [cf. Eq.~(\ref{eq:Uhat})]
\begin{eqnarray}
  \label{eq:TH_toy}
  \abs{\Theta}^2 = \frac{\abs{M_D}^2}{M_N^2}
  = \frac{M_\nu}{M_N} 
  =
  4.9 \times 10^{-11}
  \left( \frac{1\GeV}{M_N} \right)
  \left( \frac{M_\nu^2}{2.4 \times 10^{-3} \eV^2} \right)^{1/2} \,,
\end{eqnarray}
where $M_D$ and $M_N$ are the Dirac and Majorana masses, and we have
used the seesaw formula for active neutrino mass $M_\nu = |-
M_D^2/M_N|$.  Thus, the mixing is determined from the masses of active
and sterile neutrinos.

In the $\nu$MSM, since the parameter space is larger, the mixing
matrix of $N_2$ and $N_3$ is more complicated and its elements can be
much different from Eq.~(\ref{eq:TH_toy}).  Especially, as pointed out
in Ref.~\cite{Shaposhnikov:2006nn}, the larger mixing can be obtained
in the model with U(1) symmetry.  We shall reanalyze this point by
using the parametrization of the Yukawa matrix $F$ presented in
Eq.~(\ref{eq:F}).  The key for this issue is the complex parameter
$\omega$ in the $\Omega$ matrix.  It can be seen that the Yukawa
coupling constants as well as the elements of mixing matrix become
exponentially large as $F_{\alpha I}, \Theta_{\alpha I} \propto \exp
(|\mbox{Im}\omega|)$ for $\abs{\mbox{Im}\omega} \gg 1$, as long as the
seesaw approximation is valid.  It should be noted that the tiny
neutrino masses observed in the oscillation experiments can be
obtained even in this case.  To express this enhancement factor, we
introduce a parameter $X_\omega$ by
\begin{eqnarray}
  X_\omega = \exp (\mbox{Im}\omega ) \,.
\end{eqnarray}

Before going into details, let us here summarize the general
properties of mixing elements $\Theta_{\alpha I}$.
\begin{itemize}
\item[(i)] $\abs{\Theta_{\alpha I}}^2$ can be divided into 
  $X_\omega^2$, $X_\omega^0$ and $X_\omega^{-2}$
  terms.

\item[(ii)] The $X_\omega^2$ term in $\abs{\Theta_{\alpha 2}}^2M_2$ is
  exactly the same as that in $\abs{\Theta_{\alpha 3}}^2M_3$ for
  $\alpha = e, \mu, \tau$.  Similarly, the $X_\omega^{-2}$ term in
  $\abs{\Theta_{\alpha 2}}^2M_2$ is exactly the same as that in
  $\abs{\Theta_{\alpha 3}}^2M_3$.

\item[(iii)] The $X_\omega^0$ term in $\abs{\Theta_{\alpha 2}}^2M_2$
  is opposite to that in $\abs{\Theta_{\alpha 3}}^2M_3$
  for $\alpha = e, \mu, \tau$.

\item[(iv)] The coefficient of the $X_\omega^{-2}$ term in
  $\abs{\Theta_{\alpha I}}^2 M_I$ is obtained from the $X_\omega^{2}$
  term by changing $\xi \to - \xi$ for $\alpha = e, \mu, \tau$ and
  $I=2,3$.
\end{itemize}
We have confirmed these properties by direct calculations.  From now
on, we will present the expressions of $\abs{\Theta_{\alpha I}}^2$ for
$\abs{\mbox{Im} \omega} \gg 1$ and discuss how they depend on
the neutrino parameters, namely mass hierarchy, mixing angles, and CP
violating phases of active neutrinos.

We first consider $\abs{\Theta_{\alpha I}}^2$ for
$X_\omega \gg 1$ (\ie, $\mbox{Im}\omega \gg 1$) in the NH case.
The leading ${\cal O}(X_\omega^2)$ terms are found as
\begin{eqnarray}
  \label{eq:THe_NH}
  \left. \abs{\Theta_{eI}}^2 \right|_{X_\omega^2}
  \eqn{=}
  X_\omega^2 \, \frac{m_{\rm atm}}{4 M_I} \cos^2 \theta_{13}
  \left[ \tan^2 \theta_{13} + 2 \sqrt{r_m} \xi \sin (\delta + \eta)
    \sin \theta_{12} \, \tan \theta_{13} + r_m \sin^2 \theta_{12}
  \right] \,,
  \\
  \label{eq:THm_NH}
  \left. \abs{\Theta_{\mu I}}^2 \right|_{X_\omega^2}
  \eqn{=} X_\omega^2 \,
  \frac{ m_{\rm atm} }{4 M_I }  \, \sin^2 \theta_{23} \, \cos^2 \theta_{13}
  \left[ 1 + {\cal O}(\sqrt{r_m}) \right]\,,
  \\
  \label{eq:THt_NH}
  \left. \abs{\Theta_{\tau I}}^2 \right|_{X_\omega^2}
    \eqn{=} X_\omega^2 \,
  \frac{ m_{\rm atm}  }{4 M_I } 
  \, \cos^2 \theta_{23} \, \cos^2 \theta_{13}
  \left[ 1 + {\cal O}(\sqrt{r_m})\right] \,,
\end{eqnarray}
where $r_m = m_{\rm sol}/m_{\rm atm} \simeq 0.18$.  (The complete
expressions for the ${\cal O}(X_\omega^2)$ terms are collected in
App.~\ref{sec:appendix}.)  Notice that these expressions hold for both
$I=2$ and $3$ thanks to the general property (ii).  It is seen that
all the elements are proportional to $X_\omega^2 m_{\rm atm}/M_I$, and
hence they can be much larger than Eq.~(\ref{eq:TH_toy}) for $X_\omega
\gg 1$.

\begin{figure}[t]
  \centerline{
  \includegraphics[scale=1.1]{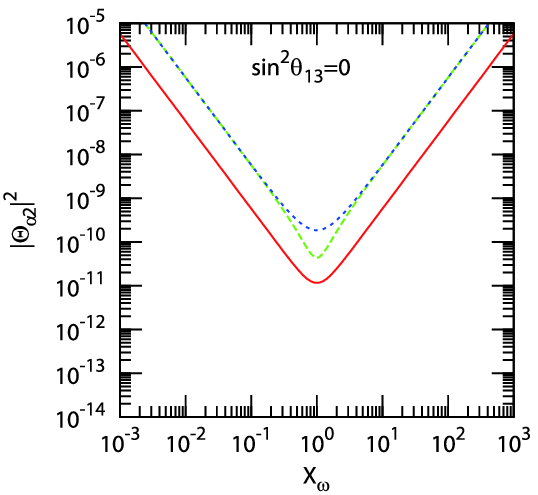}%
  \includegraphics[scale=1.1]{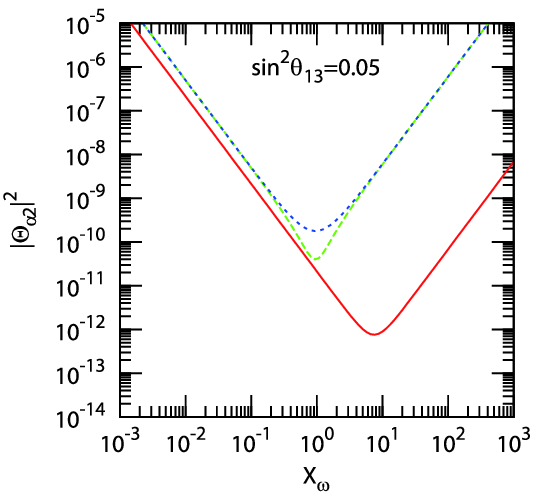}%
  \includegraphics[scale=1.1]{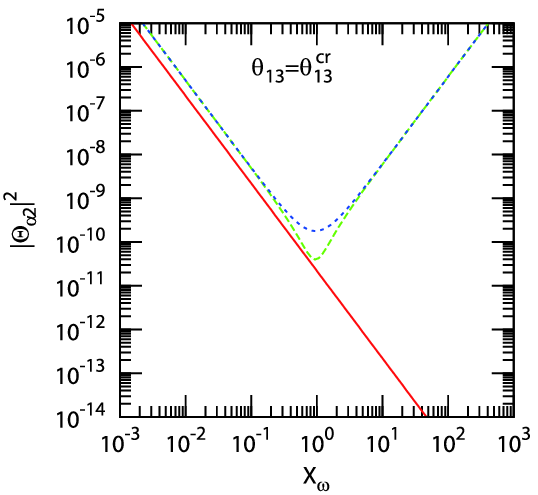}%
  }%
  \caption{ Mixing elements $|\Theta_{\alpha 2}|^2$ in the normal
    hierarchy in terms of $X_\omega$.  We take $\sin^2 \theta_{13} =0$
    (left), $0.05$ (center) and $\sin^2 \theta_{13}^{\rm
      cr}$ (right), respectively.  The red solid, green dashed
    and blue dotted lines correspond to $\abs{\Theta_{e2}}^2$,
    $\abs{\Theta_{\mu 2}}^2$ and $\abs{\Theta_{\tau 2}}^2$,
    respectively.  Here we take $M_2=120$ MeV, 
    $\mbox{Re}\omega =\pi/4$, $\delta = \pi/2$, $\eta =\pi$,
    and $\xi = +1$.  }
  \label{fig:FIG_THSQ_NH}
\end{figure}

Since the experiments show that $\theta_{13}$ is small and
$\theta_{23}$ is close to $\pi/4$, $\abs{\Theta_{\mu I}}^2$ and
$\abs{\Theta_{\tau I}}^2$ can be determined as
\begin{eqnarray}
  \label{eq:THmt_NH}
  \abs{\Theta_{\mu I}}^2 \simeq
  \abs{\Theta_{\tau I}}^2 
  \simeq 
  X_\omega^2 \frac{m_{\rm atm}}{8 M_I}
  =
  6.1 \times 10^{-12} \, X_\omega^2 \,
  \left( \frac{1\GeV}{M_I} \right) \,.
\end{eqnarray}
On the other hand, the element $\Theta_{eI}$ behaves quite
differently.  Indeed, it is interesting to note that the $X_\omega^2$
terms in $\abs{\Theta_{e2}}^2$ and $\abs{\Theta_{e3}}^2$ vanish at the
same time, when
\begin{eqnarray}
  \label{eq:CR1_NH}
  \xi \sin (\delta + \eta) = -1 \,,
\end{eqnarray}
and the mixing angle $\theta_{13}$ takes
its critical value $\theta_{13}^{\rm cr}$:
\begin{eqnarray}
  \label{eq:CRt13_NH}
  \tan \theta_{13}^{\rm cr} 
  =\sqrt{r_m} \sin \theta_{12} \,.
\end{eqnarray}
The experimental data of $\theta_{12}$, $m_{\rm atm}$ and $m_{\rm
  sol}$ with $3\sigma$ errors gives the critical value of
$\theta_{13}$ as $\sin^2 \theta_{13}^{\rm cr} = 0.041 \mbox{--}
0.070$, which can be below the $3\sigma$ upper bound $\sin^2
\theta_{13} < 0.053$.%
\footnote{When we use the data at $2\sigma$ level, $\sin^2
  \theta_{13}^{\rm cr} = 0.046\mbox{--}0.065$, which exceeds the
  $2\sigma$ bound $\sin^2 \theta_{13} < 0.039$.}  Moreover, we find in
this case that the ${\cal O}(X_\omega^0)$ term also vanishes and only
the ${\cal O}(X_\omega^{-2})$ term is left as
\begin{eqnarray}
  \label{eq:THesup_NH}
  \abs{\Theta_{e I}}^2 
  \eqn{\simeq}
  X_\omega^{-2} \frac{m_{\rm atm} r_m}{M_I}  \sin^2 \theta_{12} 
  \simeq
  2.8 \times 10^{-12} \, X_\omega^{-2}
  \left( \frac{1\GeV}{M_I} \right) \,,
\end{eqnarray}
which becomes much smaller than other elements in Eq.~(\ref{eq:THmt_NH})
for $X_\omega \gg 1$.  
We should mentioned that
the above cancellation in $\abs{\Theta_{eI}}^2$ can be realized
for any choice of masses $M_2$ and $M_3$.
As we will show in Fig.~\ref{fig:FIG_THSQ_NH},
the strong suppression in $\abs{\Theta_{eI}}^2$
is still possible when $\theta_{13}$ is close to $\theta_{13}^{\rm cr}$.
On the other hand, when $\theta_{13} =0$, 
$\abs{\Theta_{eI}}^2$ receives no strong suppression described above,
but it satisfies the relation
\begin{eqnarray}
  \label{eq:THe_NH_s13=0}
  \frac{\abs{\Theta_{eI}}^2}{\abs{\Theta_{\mu I}}^2}
  \simeq
  \frac{\abs{\Theta_{eI}}^2}{\abs{\Theta_{\tau I}}^2}
  \simeq 
  2 \, r_m \sin^2 \theta_{12}
  \simeq 0.11 \,.
\end{eqnarray}
This relation had already been obtained in
Ref.~\cite{Shaposhnikov:2008pf}.

In Fig.~\ref{fig:FIG_THSQ_NH} we show the mixing elements
$\abs{\Theta_{\alpha 2}}^2$ in terms of $X_\omega$ when $M_2 = 120$
MeV.  First, we observe that $\abs{\Theta_{\mu 2}}^2$ and
$\abs{\Theta_{\tau 2}}^2$ scales as $X_\omega^2$ for $X_\omega \gg 1$,
and they can take much larger values than the naive result in
Eq.~(\ref{eq:TH_toy}).  This behavior does not change much as long as
$\theta_{13}$ lies in the experimentally allowed region.  Second, when
$\sin^2 \theta_{13}=0$, $\abs{\Theta_{e2}}^2$ behaves similar to
$\abs{\Theta_{\mu 2}}^2$ and $\abs{\Theta_{\tau 2}}^2$ for $X_\omega
\gg 1$, but is smaller by one order of magnitude
as shown in
Eq.~(\ref{eq:THe_NH_s13=0}).  Finally, it is clearly seen that
$\abs{\Theta_{e2}}^2$ can be suppressed by many orders of magnitude
for $X_\omega \gg 1$ when $\theta_{13}$ becomes close to its
critical value with a suitable parameter choice.  Moreover,
if $\theta_{13} = \theta_{13}^{\rm cr}$, we can see that
$\abs{\Theta_{e2}}^2$ is proportional to $X_\omega^{-2}$ and it can be
extremely suppressed for $X_\omega \gg 1$.  
The cancellation of
$\abs{\Theta_{eI}}^2$ in the NH case 
is one of the most important observation in this
analysis and we will discuss its impacts on 
the experimental signatures of $N_2$
and $N_3$ later.

On the other hand, when $X_\omega \ll 1$ (\ie, $\mbox{Im}\omega \ll
-1$), the leading order contribution is proportional to
$X_\omega^{-2}$ and their expressions are given by
Eqs.(\ref{eq:THe_NH}), (\ref{eq:THm_NH}) and (\ref{eq:THt_NH}) by
replacing $X_\omega^2$ by $X_\omega^{-2}$ and $\xi$ by $- \xi$ due to
the general property (iv).  
Therefore, the above arguments with the opposite sign of $\xi$
exactly hold  and we will discard this case in the followings.

Next, we turn to discuss the IH case when $X_\omega \gg 1$.  In this
case, the leading term of the mixing element of electron type is found
as
\begin{eqnarray}
  \label{eq:THe_IH}
  \left. \abs{ \Theta_{e I} }^2 \right|_{X_\omega^2}
  \eqn{=} 
  \frac{ X_\omega^2 \, m_{2} }{4 M_I}
  \cos^2 \theta_{12} \cos^2 \theta_{13} 
  \left[ 
    \tan^2 \theta_{12} - 2 \xi \sin \eta \,
    \sqrt{\frac{m_1}{m_2}} \tan \theta_{12}
    + \frac{m_1}{m_2}
  \right] \,,
\end{eqnarray}
and the expressions for $\abs{\Theta_{\mu I}}^2$ and
$\abs{\Theta_{\tau I}}^2$ are so long and they are collected in
App.~\ref{sec:appendix}.  It is interesting to note that the
$X_\omega^2$ term as well as the $X_\omega^0$ term in
$\abs{\Theta_{eI}}^2$ vanishes when
\begin{eqnarray}
  \xi \sin \eta = + 1 ~~
  \mbox{and}~~
  \tan \theta_{12} =
  \tan \theta_{12}^{\rm cr} =
  \sqrt{ \frac{m_1}{m_2} } = (1+ r_m^2)^{-1/4} \,,
\end{eqnarray}
and then $\theta_{12}^{\rm cr}$
is close to the maximal angle $\pi/4$.
Unfortunately, 
it is far beyond the current data of 
$3\sigma$ range, and the cancellation in $\abs{\Theta_{eI}}^2$
cannot be realized in the IH case, which is different from
the NH case.%
\footnote{
When $\xi \sin \eta = -1$, $\tan \theta_{12} = \sqrt{m_2/m_1}$,
and $\sin \theta_{13} =0$, 
the $X_\omega^2$ terms in both $\abs{\Theta_{\mu I}}^2$
and $\abs{\Theta_{\tau I}}^2$ vanish at the same time.
However, the required value of $\theta_{12}$ is not 
allowed by the current data.}
This point gives significant effects
on the discussions given in the following sections.

\begin{figure}[t]
  \centerline{
  \includegraphics[scale=1.1]{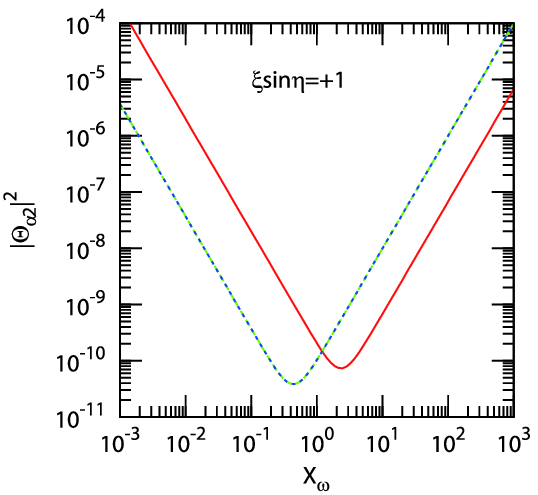}
  \includegraphics[scale=1.1]{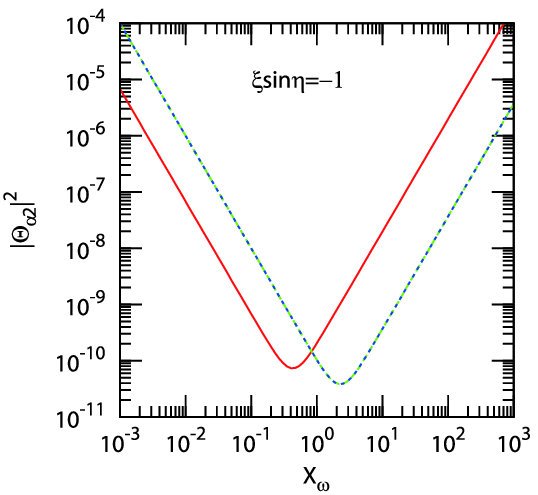}
  }%
  \caption{ 
    Mixing elements $|\Theta_{\alpha 2}|^2$
    in the inverted hierarchy in terms of $X_\omega$. 
    We take $\sin \eta =\pi/2$ (left)
    and $3 \pi/2$ (right), respectively.
    The red solid, green dashed and blue dotted lines
    correspond to $\abs{\Theta_{e2}}^2$, $\abs{\Theta_{\mu 2}}^2$
    and $\abs{\Theta_{\tau 2}}^2$, respectively.
    Here we take $M_2=120$ MeV, 
    $\mbox{Re}\omega =\pi/4$,
    $\delta = \pi/2$, $\theta_{13} =0$,
    and $\xi = + 1$.
  }
  \label{fig:FIG_THSQ_IH}
\end{figure}

However,
``$\xi \sin \eta$'' plays a crucial 
role to determine the mixing element $\abs{\Theta_{eI}}$.
To see this point, let us take 
$\theta_{23} = \pi/4$ and $\theta_{13} = 0$ for simplicity.
In this case, the mixing elements are given by
\begin{eqnarray}
  \left. \abs{\Theta_{eI}}^2 \right|_{X_\omega^2}
  \eqn{\simeq} \frac{X_\omega^2 m_{\rm atm}}{4 M_I}
  ( 1 - \xi \sin \eta \sin 2 \theta_{12} ) \,,
  \\
  \left. \abs{\Theta_{\mu I}}^2 \right|_{X_\omega^2}
  \eqn{\simeq} 
  \left. \abs{\Theta_{\tau I}}^2 \right|_{X_\omega^2}
  \simeq
  \frac{X_\omega^2 m_{\rm atm}}{8 M_I}
  ( 1 + \xi \sin \eta \sin 2 \theta_{12} ) \,,
\end{eqnarray}
which gives the relation~\cite{Shaposhnikov:2008pf}
\begin{eqnarray}
  \label{eq:Ratios_IH}
  \frac{ \abs{\Theta_{eI}}^2}{ \abs{\Theta_{\mu I}}^2}
  \simeq
  \frac{ \abs{\Theta_{eI}}^2}{ \abs{\Theta_{\tau I}}^2}
  \simeq
  2 \, \frac{1 - \xi \sin \eta \sin 2 \theta_{12}}
  {1 +\xi \sin \eta \sin 2 \theta_{12}}
  =
  \left\{
    \begin{array}{l l}
      0.071 &~~\mbox{for}~\xi \sin\eta = +1 \\
      56    &~~\mbox{for}~\xi \sin\eta = -1 
    \end{array}
  \right. \,.
\end{eqnarray}
Therefore, the mixing element of electron type can be smaller or
larger than others depending on the choice of ``$\xi \sin \eta$''.
This property is represented in Fig.~\ref{fig:FIG_THSQ_IH}.  It is
seen that $\abs{\Theta_{\mu I}}^2$ and $\abs{\Theta_{\tau I}}^2$ are
almost the same, but $\abs{\Theta_{eI}}^2$ can be different from
others.

Before closing this section, we would like to stress again that the
above results of the hierarchy between the mixing elements
$\Theta_{\alpha I}$ are independent on the masses of sterile
neutrinos.  Therefore, they can be applied to the general seesaw
models with two right-handed neutrinos.

\section{Neutrinoless Double Beta Decay}
\label{sec:0nu2beta}
The neutrinoless double beta ($0 \nu 2 \beta$) decay is one important
phenomenon in which the mixing of active and sterile neutrinos,
$U_{\alpha i}$ and $\Theta_{\alpha I}$, plays a crucial role.  The $0
\nu 2 \beta$ decay in the $\nu$MSM had already been investigated in
Ref.~\cite{Bezrukov:2005mx}, in which the contributions from active
neutrinos and dark-matter sterile neutrino $N_1$ are estimated and
those from $N_{2,3}$ are neglected since their masses are assumed to
be so heavy that they decouple from the considering decay processes.
Further, it had been discussed in Ref.~\cite{Benes:2005hn} that the $0
\nu 2 \beta$ decay gives the stringent constraint on the mixing
element $\Theta_{eI}$ especially when their masses are around 100 MeV.
Following to these analyses, we would like to reconsider this issue in
this section.  In particular, we shall take into account the
contributions from $N_{2,3}$ by considering a wider range of their
masses, and also the mixing elements $\Theta_{\alpha 2,3}$ discussed
in the previous section.  We will then show that the constraints from
the $0 \nu 2 \beta$ decay are negligible in the $\nu$MSM.

The rate of the $0\nu 2\beta$ decay is characterized by the effective
neutrino mass $m_{\rm eff}$ (see, \eg, Ref.~\cite{0nu2beta}).  In the
$\nu$MSM it is given by
\begin{eqnarray}
  \label{eq:meff_nuMSM}
  m_{\rm eff} =
  m_{\rm eff}^\nu
  + 
  \sum_{I=1,2,3} M_I \, \Theta_{e I}^2 \,
  f_\beta(M_I) 
  \,,
\end{eqnarray}
where the first term $m_{\rm eff}^\nu$ denotes the contribution from
active neutrinos
\begin{eqnarray}
  m_{\rm eff}^\nu = \sum_{i=1,2,3} m_i U_{ei}^2 \,.
\end{eqnarray}
The second term in Eq.~(\ref{eq:meff_nuMSM}) denotes the contribution
from sterile neutrinos in which we have introduced the function
$f_\beta$ to represent the suppression of the nuclear matrix element
from neutrinos with masses heavier than about
100~MeV~\cite{Hirsch:1995rf,Blennow:2010th}.  In this analysis, for
simplicity, we shall assume $f_{\beta}(M_I) = 1$ for $M_I \le
\Lambda_{\beta}$ and $f_{\beta}(M_I) = (\Lambda_{\beta}/M_I)^2$ for
$M_I > \Lambda_{\beta}$ where the typical energy scale in the matrix
element is taken as $\Lambda_\beta = 100$ MeV.  The more precise
treatment of the function $f_\beta$ at $M_I \simeq \Lambda_\beta$ does
not alter our final conclusions.

First, we consider the case when all the sterile neutrinos in the
$\nu$MSM are lighter than $\Lambda_\beta$.  As for the dark-matter
sterile neutrino $N_1$, its mass is indeed smaller than
$\Lambda_\beta$.  On the other hand, the masses of $N_2$ and $N_3$ can
be smaller or larger than $\Lambda_\beta$.  Here we consider the
former case and the latter case will be discussed separately below.
In this case the effective neutrino mass exactly vanishes.  This is
because
\begin{eqnarray}
  \label{eq:meff_0}
  m_{\rm eff} 
  = \sum_{i=1,2,3} m_i \, U_{e i}^2 
  + 
  \sum_{I=1,2,3} M_I \, \Theta_{e I}^2 
  =
  [ \, \hat U \, \hat M_\nu^{\rm diag} \, \hat U^T \, ]_{ee}
  = [\hat M_\nu]_{ee} = 0 \,.
\end{eqnarray}
This cancellation has been recently observed in
Ref.~\cite{Blennow:2010th} by using the general seesaw model.  Thus,
the $\nu$MSM with $M_{1,2,3} < \Lambda_{\beta}$ predicts zero event in
the $0\nu 2 \beta$ decay, which means that
the current experimental limits on $m_{\rm eff}$ give
no constraint on the model.

On the other hand, $N_{2,3}$ can be heavier than $\Lambda_\beta$ in
the $\nu$MSM.  In this case, the prediction of $m_{\rm eff}$ is
modified.  We consider the case when $N_1$ is resonantly produced via
thermal scatterings in the presence of the lepton asymmetries as
mentioned in Sec.~2.  In this case, as already pointed out in
Ref.~\cite{Bezrukov:2005mx}, the contribution from dark-matter sterile
neutrino $N_1$, denoted by $m_{\rm eff}^{N_1}$, is very small.  The
correct abundance of dark matter can be obtained when $M_1 =4$--50~keV
and $|F_{\alpha 1}| \simeq 5 \times 10^{-15}$--$4 \times
10^{-13}$~\cite{Laine:2008pg}.  Since $M_1 < \Lambda_{\beta}$, we find
\begin{eqnarray}
  |m_{\rm eff}^{N_1}| = M_1 \, |\Theta_{e1}^2|
  = \frac{|F_{e1}^2| \, \vev{\Phi}^2}{M_1}
  = {\cal O}(10^{-11} \mbox{--}10^{-6}) \, \eV \,,
  \label{eq:meff_N1}
\end{eqnarray}
which is so small to be negligible in most cases, but we will point
out its importance.

Let us turn to consider the contribution from $N_2$ and $N_3$.  These
sterile neutrinos are quasi-degenerate in order for the successful
baryogenesis~\cite{Akhmedov:1998qx,Asaka:2005pn,Asaka:2010kk,Canetti:2010aw}.
We then write the masses as $M_3 = M_N + \Delta M/2$ and $M_2 = M_N -
\Delta M/2$ with $\Delta M \ll M_N$, and divide $m_{\rm eff}$ from
$N_{2,3}$ into two parts
\begin{eqnarray}
  m_{\rm eff}^{N_{2,3}} = 
  \sum_{I=2,3} M_I \Theta_{eI}^2 f_\beta (M_I) 
  = \bar{m}_{\rm eff}^{N_{2,3}} + \delta m_{\rm eff}^{N_{2,3}} \,,
\end{eqnarray}
where 
\begin{eqnarray}
  \bar{m}_{\rm eff}^{N_{2,3}} 
  \eqn{=} 
  f_\beta (M_N) \sum_{I=2,3} M_I \Theta_{eI}^2  \,,
  \\
  \delta m_{\rm eff}^{N_{2,3}}
  \eqn{=}
  \sum_{I=2,3} 
  \left[ f_\beta (M_I) - f_\beta(M_N) \right] \, M_I
  \Theta_{eI}^2 \,.
\end{eqnarray}
Notice that the second part $\delta m_{\rm eff}^{N_{2,3}}$ vanishes
when $\Delta M = 0$, and more interestingly, that the
first term can be written as
\begin{eqnarray}
  \bar{m}_{\rm eff}^{N_{2,3}} = 
  - f_\beta (M_N) \, m_{\rm eff}^\nu \,.
\end{eqnarray}
Therefore, the effective neutrino mass in the $\nu$MSM
when $M_{2,3} > \Lambda_\beta$ is written as
\begin{eqnarray}
  \label{eq:meff_nuMSM1}
  m_{\rm eff} = \left[ 1 - f_\beta (M_N) \right] m_{\rm eff}^{\nu}
  + m_{\rm eff}^{N_1} + \delta m_{\rm eff}^{N_{2,3}} \,.
\end{eqnarray}

\begin{figure}[t]
  \centerline{
    \includegraphics[scale=1]{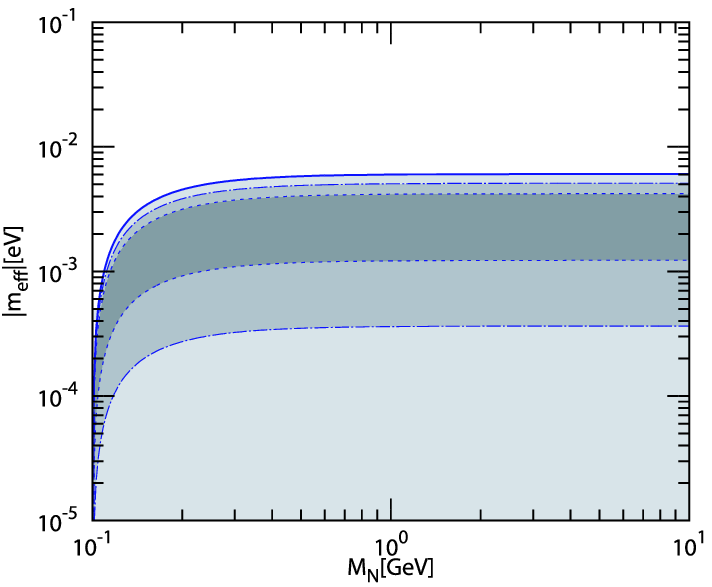}\hspace{5ex}
    \includegraphics[scale=1]{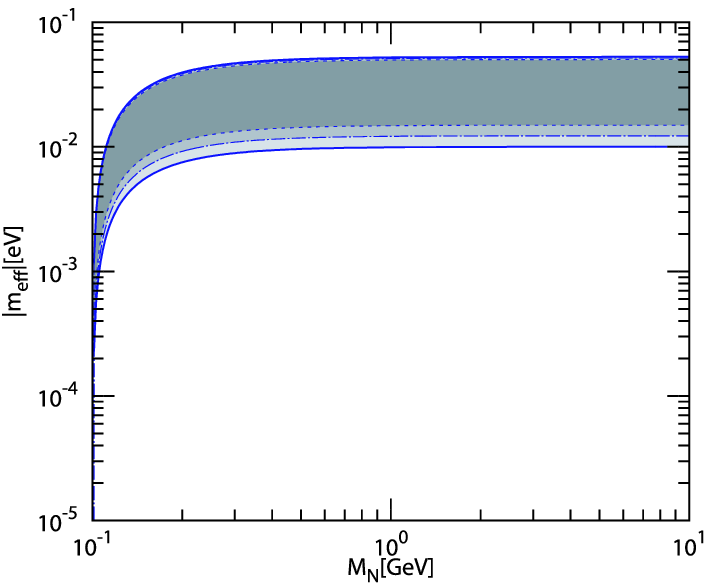}
  }%
  \caption{Allowed regions for $|m_{\text{eff}}|$ as a function
          of the mass of sterile neutrino $M_N$, for NH (left)
          and IH (right) cases, respectively.
          The light gray region with solid lines, 
          gray region with dot-dashed lines, 
          and dark gray region with dotted lines
          correspond to regions allowed 
          by the observational data for active neutrinos 
          at the 3$\sigma$, 2$\sigma$ and 1$\sigma$ level, 
          respectively.
          Here we take $\Delta M / M_{N}=5 \times 10^{-8}$.
  }
 \label{fig:meff}
\end{figure}

It is then found that when $M_N \gg \Lambda_\beta$ (and hence $f_\beta
(M_{2,3}) \ll 1$)
\begin{eqnarray}
  m_{\rm eff} \simeq m_{\rm eff}^\nu + m_{\rm eff}^{N_1} \,,
\end{eqnarray}
which shows the sizable contributions come only from active neutrinos
and dark-matter sterile neutrino~\cite{Bezrukov:2005mx}.  On the other
hand, $N_2$ and $N_3$ give a significant effect on $m_{\rm eff}$
especially when they are quasi-degenerate and $M_N \simeq
\Lambda_\beta$, namely, they induce the destructive contribution
to $m_{\rm eff}$ given in Eq.~(\ref{eq:meff_nuMSM}).  Thus,
$|m_{\rm eff}|$ in the $\nu$MSM can be much smaller than $|m_{\rm
  eff}^\nu|$ when $M_N \simeq \Lambda_\beta$.
This point is illustrated in Fig.~\ref{fig:meff}, 
where we show the allowed region of $|m_{\rm eff}|$
in terms of $M_N$ by varying 
$\mbox{Re}\omega$, $\delta$ and $\eta$ in the range
$[0, 2\pi]$,  $\mbox{Im}\omega$ in the range $[0,7]$,
and the parameters
of active neutrinos within the experimental 3$\sigma$ range.

Next, we turn to see how $m_{\rm eff}$ depends on masses, mixing
angles and CP violating phases of active neutrinos.  In the NH case,
the contribution from active neutrinos is (\eg, see
\cite{Aalseth:2004hb})
\begin{eqnarray}
  m_{\rm eff}^\nu
  =
  e^{-2i\delta}
  \left(
    \sin^2 \theta_{13} \, m_3 +
    e^{2i(\delta + \eta)} \, \cos^2 \theta_{13} \, 
    \sin^2 \theta_{12} \, m_2
  \right) \,.
\end{eqnarray}
Notice that $m_1 = 0$ in the limit of $F_{\alpha 1} \to 0$.
We also find that
\begin{eqnarray}
  \delta m_{\rm eff}^{N_{2,3}}
  \eqn{=}
  \frac{\Delta M}{2} f_\beta'(M_N)
  \left[
    M_3 \, \Theta_{e3}^2 - M_2 \, \Theta_{e2}^2
  \right] 
  \nonumber \\
  \eqn{=}
  e^{-2 i \delta}
  \frac{\Delta M \, \Lambda_\beta^2}{2 M_N^3} 
  \Bigl[
  X_\omega^2 \, e^{-2i {\rm Re}\omega}
  \left( 
    \sqrt{m_3}\xi \sin \theta_{13}
    - i e^{i(\delta+\eta)} \sqrt{m_2}
    \cos \theta_{13} \sin \theta_{12}
  \right)^2
  \nonumber \\
  \eqn{}~~~~~~~~~~~~~~
  +
  X_\omega^{-2} \, e^{+2i {\rm Re}\omega}
  \left( 
    \sqrt{m_3}\xi \sin \theta_{13}
    + i e^{i(\delta+\eta)} \sqrt{m_2}
    \cos \theta_{13} \sin \theta_{12}
  \right)^2
  \Bigr] \,,
\end{eqnarray}
where we have neglected the higher order terms of $\Delta M$.

It should be noted that $m_{\rm eff}^\nu$ and also the $X_\omega^2$
term in $\delta m_{\rm eff}^{N_{2,3}}$ vanish, if $\cos 2(\delta
+\eta) = -1$ and $\tan^2 \theta_{13} = r_m \sin^2 \theta_{12}$.
Interestingly, the required conditions are the same
as~(\ref{eq:CR1_NH}) and (\ref{eq:CRt13_NH}), \ie, those for the
cancellation in $|\Theta_{eI}|^2$.  In this case, $\delta m_{\rm
  eff}^{N_{2,3}}$ becomes suppressed for a large $X_\omega$ together
with a small mass difference of $N_{2,3}$.  As a result, the
dark-matter sterile neutrino gives the dominant contribution to the
effective mass in the $0 \nu 2 \beta$ decay $|m_{\rm eff}| \simeq
|m_{\rm eff}^{N_1}|$, which is very small as shown in
Eq.~(\ref{eq:meff_N1}).  This is the reason why the $3\sigma$ lower
bound on $|m_{\rm eff}|$ is beyond the region of the plot.

When the conditions~(\ref{eq:CR1_NH}) and (\ref{eq:CRt13_NH}) are not
satisfied, we find for large $X_\omega$
\begin{eqnarray}
  |\delta m_{\rm eff}^{N_{2,3}} |
  \simeq 1.5 \times 10^{-12} \eV 
  \left( \frac{\Delta M/M_N}{5 \times 10^{-8}} \right)
  \left( \frac{1 \, \GeV}{M_N} \right)^2 X_\omega^2
  \,,
\end{eqnarray}
by using the central values of parameters of active neutrinos and
$\delta + \eta = \pi/2$.%
\footnote{ Here the maximal value of $|\delta m_{\rm eff}^{N_{2,3}} |$
  is shown.  When $\delta + \eta = 3\pi/2$, it takes the minimal value
  and the prefactor becomes $1.8 \times 10^{-13}$.}
This shows that we may neglect $\delta m_{\rm eff}^{N_{2,3}}$ as long
as $\Delta M$ is sufficiently small.  Note that $X_\omega$ cannot be
so large due to the experimental upper bounds on $\Theta_{\alpha I}$,
which will be investigated in the next section.  In this case, the
effective neutrino mass becomes
\begin{eqnarray}
  m_{\rm eff} \simeq \left[ 1 - f_\beta (M_N) \right] m_{\rm eff}^{\nu}
  + m_{\rm eff}^{N_1} \,.
\end{eqnarray}
We then understand the $3\sigma$ upper bound on $|m_{\rm eff}|$ in
Fig.~\ref{fig:meff} as $\left[ 1 - f_\beta (M_N) \right] m_{\rm
  eff}^{\nu} \bigr|_{\rm MAX}$, which becomes suppressed when $M_N$ is
close to $\Lambda_\beta$.

In the IH case, we obtain
\begin{eqnarray}
  m_{\rm eff}^{\nu}
  =
  \cos^2 \theta_{13}
  \left(
    \cos^2 \theta_{12} \, m_1 +
    e^{2i \eta} \, \sin^2 \theta_{12} \, m_2
  \right) \,,
\end{eqnarray}
and
\begin{eqnarray}
  \delta m_{\text{eff}}^{N_{2,3}} 
  \eqn{=}
  \frac{\Delta M}{2} f_\beta^{\prime}(M_N) 
  \left[ M_3 \, \Theta_{e3}^2 - M_2 \, \Theta_{e2}^2 \right] 
  \nonumber \\
  \eqn{=} - \frac{\Delta M \Lambda_\beta^2 }{2 M_N^3} 
  \cos^2{\theta_{13}} 
  \left[ X_{\omega}^2 \, e^{-2 i \mbox{Re}\omega} 
    \left( \sqrt{m_1} \xi \cos{\theta_{12}} 
      + i \, e^{i \eta} \sqrt{m_2} \sin{\theta_{12}} 
    \right)^2 \right. 
  \nonumber \\
  &&\hspace{80pt} + \left. 
    X_{\omega}^{-2} \, e^{+2 i \mbox{Re}\omega} 
    \left( \sqrt{m_1} \xi \cos{\theta_{12}} 
      - i \, e^{i \eta} \sqrt{m_2} \sin{\theta_{12}} \right)^2 
  \right] \,.
\end{eqnarray}
Then, as in the NH case, $m_{\rm eff}^\nu$ as well as the $X_\omega^2$
term in $\delta m_{\rm eff}^{N_{2,3}}$ vanish, when $\xi \sin \eta =
+1$ and $\tan^2 \theta_{12} = m_1/m_2$.  The required conditions are
the same as those of the cancellation in $\abs{\Theta_{eI}}$, and they
are not satisfied the present experimental data on the masses and
mixing angles of active neutrinos.  This is the reason why $|m_{\rm
  eff}|$ in the IH case receives the stringent lower bound as shown in
Fig.~\ref{fig:meff}, and $m_{\rm eff}^{N_1}$ becomes negligible.
Further, compared with other terms, $\delta m_{\rm eff}^{N_{2,3}}$ can
be neglected for a small mass difference of $N_{2,3}$.
This is because
\begin{eqnarray}
  |\delta m_{\rm eff}^{N_{2,3}} |
  \simeq 2.3 \times 10^{-11} \eV 
  \left( \frac{\Delta M/M_N}{5 \times 10^{-8}} \right)
  \left( \frac{1 \, \GeV}{M_N} \right)^3 X_\omega^2 \,,
\end{eqnarray}
when we take $\eta = 3 \pi/2$.
\footnote{Here the maximal value of $|\delta m_{\rm eff}^{N_{2,3}} |$
  is shown.  When $\eta = \pi/2$, it takes 
  the minimal value and the prefactor becomes $8.0 \times
  10^{-13}$.}
Therefore, we obtain in the IH case
\begin{eqnarray}
  m_{\rm eff} \simeq \left[ 1 - f_\beta (M_N) \right] m_{\rm eff}^{\nu} \,.
\end{eqnarray}

In summary, we have shown that the effective neutrino mass in the $0
\nu 2 \beta$ decay is given by the contribution from active neutrinos
$m_{\rm eff}^\nu$ (together with a very small $m_{\rm eff}^{N_1}$) for
$M_{2,3} \gg \Lambda_\beta$.  On the other hand, when $M_{2,3} \sim
\Lambda_\beta$, sterile neutrinos $N_2$ and $N_3$ give
the destructive contribution and then $m_{\rm eff}$ becomes much smaller
than that from active neutrinos.  Especially, $m_{\rm eff}$ vanishes
when all the sterile neutrinos are lighter than $\Lambda_\beta$.
Therefore, the $\nu$MSM receives no stringent constraint
on the mixing elements $\Theta_{\alpha I}$ from the $0 \nu 2 \beta$ 
decay pointed out in Ref.~\cite{Benes:2005hn}.

\section{Search for light sterile neutrinos}
\label{sec:search}
In this section we shall discuss the experimental search of
sterile neutrinos $N_{2,3}$.  In particular, we restrict ourselves
here to the case when the masses are smaller than the pion mass.
This is simply because we would like to discuss the production
of $N_{2,3}$ in the decays of $\pi^+$ and $K^+$.  In addition,
although such light sterile neutrinos have a long lifetime and may
decay after the big bang nucleosynthesis (BBN) starts as we will show
below, it can be avoided the stringent constraint from the
BBN~\cite{Kawasaki:2004qu} on the hadronic decays of $N_{2,3}$.  In
this section, we shall explore in detail the impacts of the results in
Sec.~\ref{sec:mixing} on the direct searches of $N_{2,3}$ 
in the $\pi^+$ and $K^+$ decays.

\begin{figure}[t]
  \centerline{
  \includegraphics[scale=1.3]{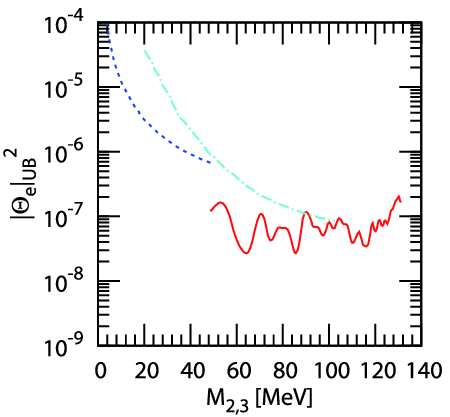}%
  \includegraphics[scale=1.3]{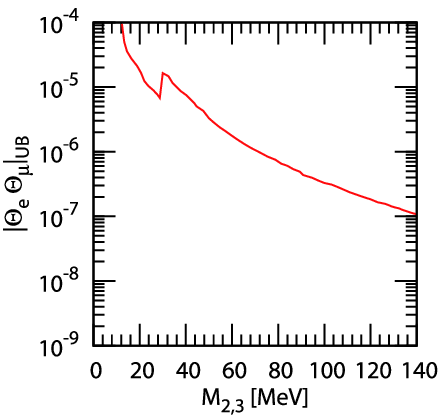}%
  \includegraphics[scale=1.3]{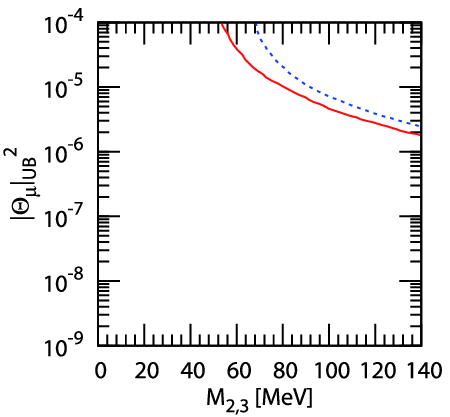}%
  }%
  \caption{ Upper bounds on $|\Theta_{eI}|^2$ (left), $|\Theta_{eI}
    \Theta_{\mu I}|$ (center) and $|\Theta_{\mu I}|^2$ (right) from
    direct searches of sterile neutrino.  In the left panel the red
    sold line is from \cite{Britton:1992xv}, the blue dotted line is
    from \cite{Britton:1992pg}, and the cyan dot-dashed line is from
    \cite{Bernardi:1987ek}.  In the center panel the red solid line is
    from \cite{Bernardi:1987ek}.  In the right panel 
    the red solid line is from \cite{Yamazaki:1984}, and
    the blue dashed line is from \cite{Hayano:1982wu}.
  }
  \label{fig:FIG_THSQUB}
\end{figure}

There have been so far various experiments of the direct search for
sterile neutrino which give the upper bounds on the mixing elements
$\Theta_{\alpha I}$.  In the considering mass region, the signal of
sterile neutrino can be investigated by the peak
search~\cite{Shrock:1980vy} in the energy spectrum of charged leptons
from meson decays~\cite{Britton:1992xv}-\cite{Hayano:1982wu}, and also
by finding charged leptons from the decays of sterile neutrino inside
the detector~\cite{Bernardi:1985ny}-\cite{Orloff:2002de}.  See, for
example, Refs.~\cite{Kusenko:2004qc,Gorbunov:2007ak,Atre:2009rg}.  The
upper bounds on the mixing elements are summarized in
Fig.~\ref{fig:FIG_THSQUB}.  Notice that the upper bound on
$|\Theta_{\tau I}|$~\cite{Orloff:2002de} is weaker than that on
$|\Theta_{\mu I}|$, and then it plays no significant role for our
conclusion.  It can be seen that the bound on $|\Theta_{eI}|$ is
severer than $|\Theta_{\mu I}|$ by orders of magnitude.

\begin{figure}[tb]
  \centerline{
    \includegraphics[scale=1]{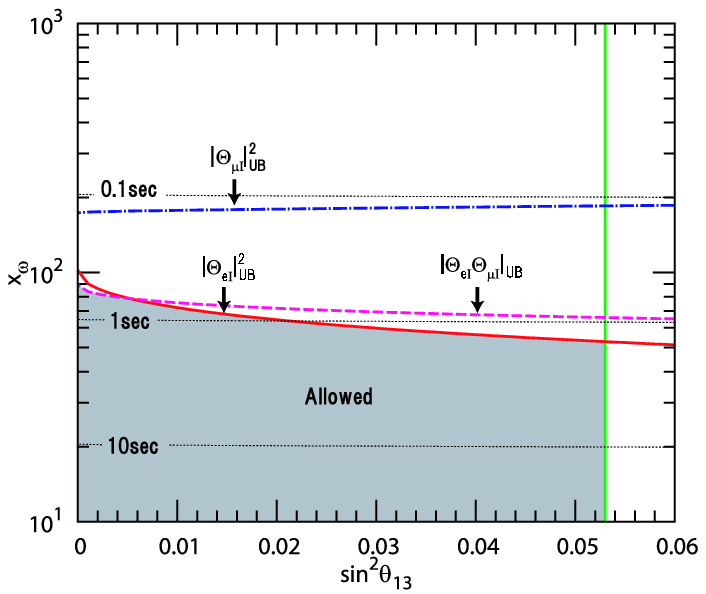}%
    \hspace{2ex}
    \includegraphics[scale=1]{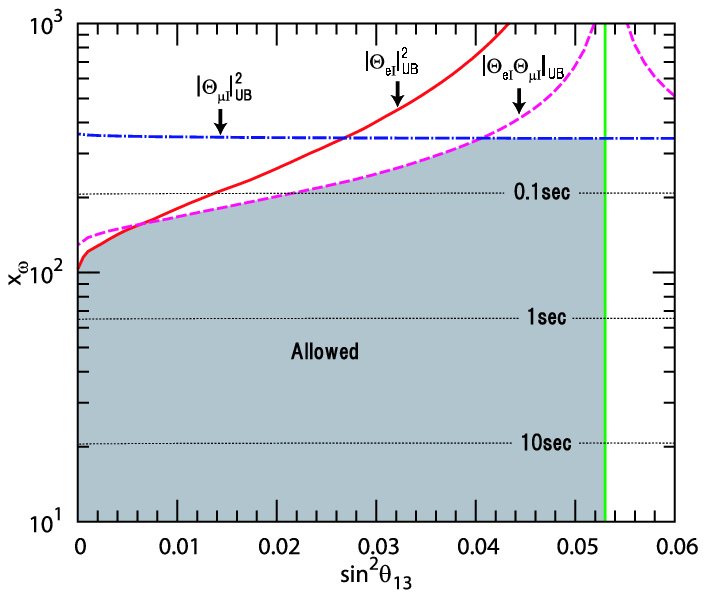}
  }
  \caption{Allowed region in the
    $\sin^2{\theta_{13}}\text{-}X_{\omega}$ plane for the NH case by
    taking $\delta = 0$ and $\eta = \frac{\pi}{2}$ (left) and $\delta
    = 0$ and $\eta = \frac{3\pi}{2}$ (right).  The red solid, the
    magenta dashed, and the blue dot-dashed lines are the upper bounds
    on $X_{\omega}$ from the experimental limits on $|\Theta_{e
      I}|^2$, $|\Theta_{e I} \Theta_{\mu I}|$, and $|\Theta_{\mu
      I}|^2$ respectively.  The green line is the $3 \sigma$ limit on
    $\sin^2 \theta_{13}$.  The black dotted lines show lifetimes of
    sterile neutrino.  We take $M_{3} = 120$MeV, $\Delta M^2/M_{3}^2 =
    10^{-8}$, Re$\omega = \frac{\pi}{4}$, and $\xi = + 1$.}
  \label{fig:CONTOUR-NH}
\end{figure}
%
\begin{figure}[!tb]
  \centerline{
    \includegraphics[scale=1]{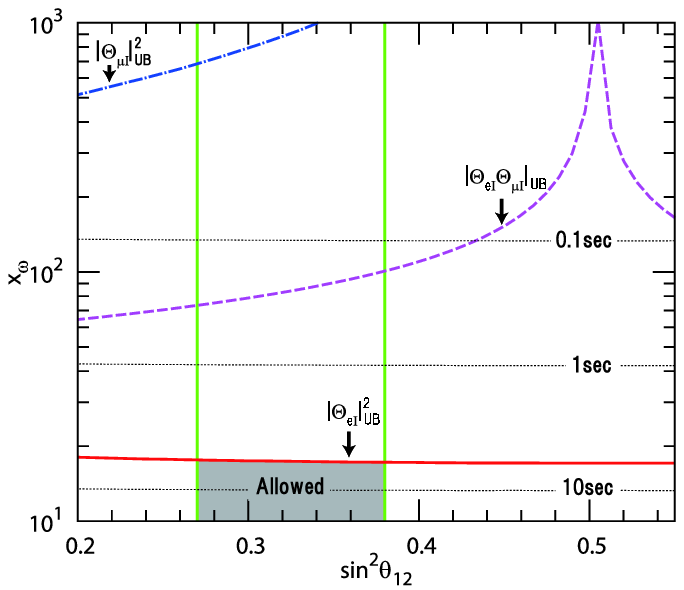}
    \hspace{2ex}
    \includegraphics[scale=1]{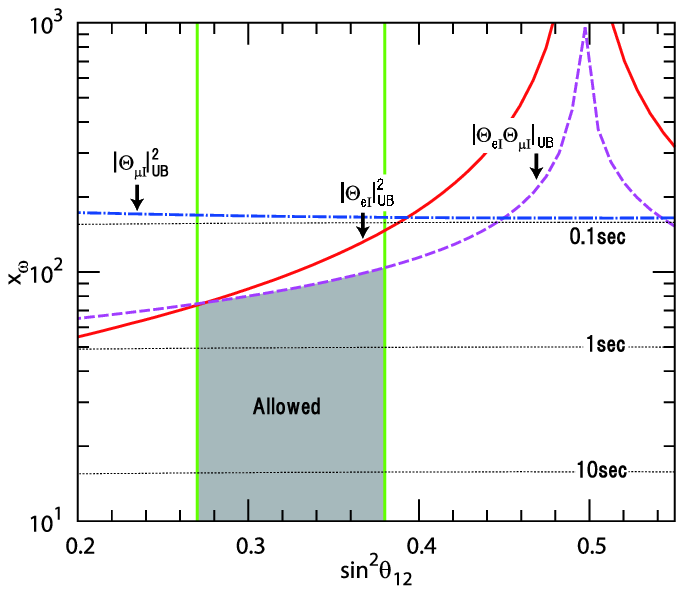}
  }
  \caption{Allowed region in the
    $\sin^2{\theta_{12}}\text{-}X_{\omega}$ plane for the IH case by
    taking $\delta = \frac{\pi}{2}$ and $\eta = \frac{3\pi}{2}$ (left)
    and $\delta = \frac{\pi}{2}$ and $\eta = \frac{\pi}{2}$ (right).
    The red solid, the magenta dashed, and the blue dot-dashed lines
    are the upper bounds on $X_{\omega}$ from the experimental limits
    on $|\Theta_{e I}|^2$, $|\Theta_{e I} \Theta_{\mu I}|$, and
    $|\Theta_{\mu I}|^2$ respectively.  The green lines shows the $3
    \sigma$ range of $\sin^2 \theta_{12}$.  The black dotted lines
    show lifetimes of sterile neutrino.  We take $M_{3} = 120$MeV,
    $\Delta M^2/M_{3}^2 = 10^{-8}$, $\theta_{13}=0$, 
    Re$\omega = \frac{\pi}{4}$ and $\xi = +1$. }
 \label{fig:CONTOUR-IH}
\end{figure}
The mixing elements of $N_{2,3}$ scale as $|\Theta_{\alpha I}| \propto
X_\omega$ for $X_\omega \gg 1$, as described in Sec.~\ref{sec:mixing}.
Thus, the experimental upper bounds on $|\Theta_{\alpha I}|$ can be
translated into the upper bound on $X_\omega$.  Such a bound is the
basis of finding the shortest lifetimes of $N_2$ and $N_3$ and also the
largest production rates in the decays of $\pi^+$ and $K^+$.  Let us
first discuss the case when $M_N = 120$ MeV.  In this case, the
experimental bounds are $|\Theta_{e I}|^2_{\rm UB} =
6.0\times10^{-8}$, $|\Theta_{e I} \Theta_{\mu I}|_{\rm UB} =
1.8\times10^{-7}$, and $|\Theta_{\mu I}|^2_{\rm UB} =
2.8\times10^{-6}$.

In the NH case, we show in Fig.~\ref{fig:CONTOUR-NH} the upper bounds
on $X_\omega$ in terms of $\sin^2 \theta_{13}$ since the element
$\Theta_{eI}$ is crucially dependent on this mixing angle.  When there
is no cancellation in $\Theta_{eI}$ (see the left panel of
Fig.~\ref{fig:CONTOUR-NH}), the experimental bound on
$\abs{\Theta_{eI}}^2$ determines the upper bound on $X_\omega$ in most
cases.  Note that, when $\theta_{13}$ is close to zero, $|\Theta_{eI}|^2$
is smaller than $|\Theta_{\mu I}|^2$ by one order of magnitude as
shown in Eq.~(\ref{eq:THe_NH_s13=0}) and then 
$\abs{\Theta_{eI} \Theta_{\mu I}}_{\rm UB}$ puts the upper bound on
$X_\omega$.

\begin{figure}[tb]
 \begin{minipage}{0.5\hsize}
   \centering
    \includegraphics[scale=1]{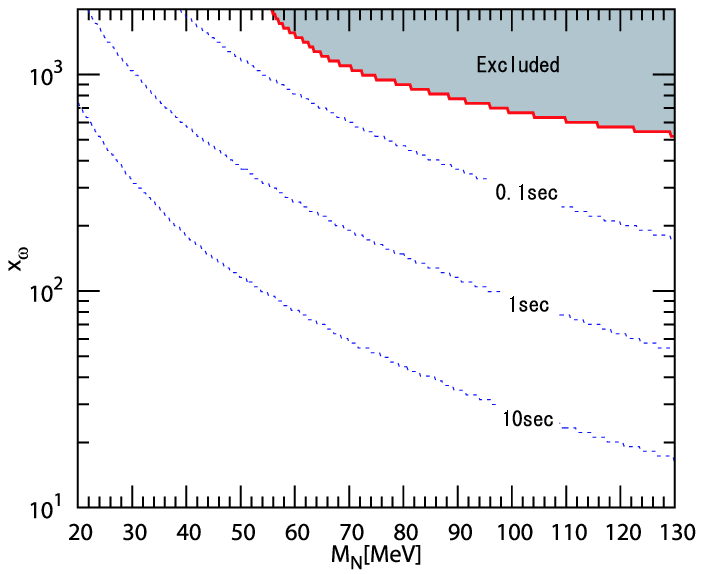}\\
 \end{minipage}
 \begin{minipage}{0.5\hsize}
   \centering
    \includegraphics[scale=1]{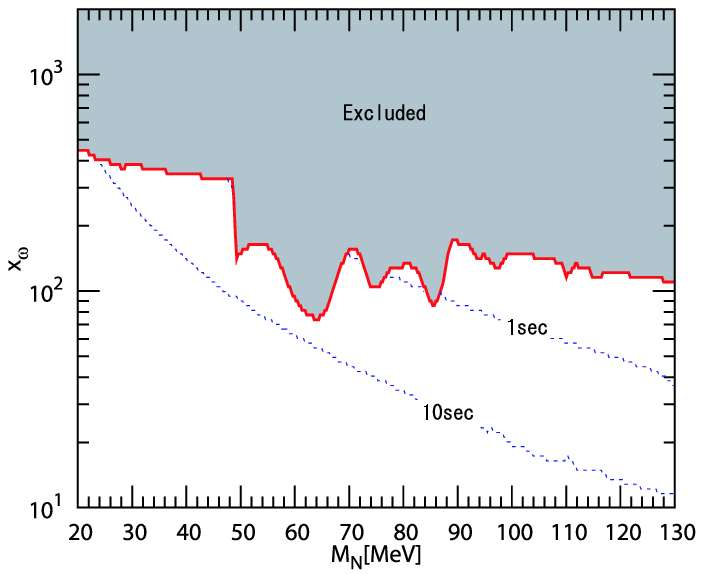}\\
 \end{minipage}
 \caption{Allowed region in the $M_{N}\text{-}X_{\omega}$ plane for the
          NH (left) and IH (right) cases.
          The gray region is excluded by experiments for direct search
          of sterile neutrinos. 
          The dotted lines show the lower bounds on $X_\omega$ for given lifetime of sterile neutrino.
          We use the observational data for active neutrinos 
          at the 3$\sigma$ level.}
 \label{fig:AR}
\end{figure}

On the other hand, as shown in the right panel of
Fig.~\ref{fig:CONTOUR-NH}, the situation is drastically changed when
the cancellation in $\Theta_{eI}$ can happen as shown in
Eqs.~(\ref{eq:CR1_NH}) and (\ref{eq:CRt13_NH}).  In this case, the
experimental bounds on $\abs{\Theta_{eI}}^2$ and also
$\abs{\Theta_{eI} \Theta_{\mu I}}$ play no significant roles, 
but that on $\abs{\Theta_{\mu I}}^2$
determines the upper bound on $X_\omega$.  Since the present 
limit $\abs{\Theta_{\mu I}}^2_{\rm UB}$ is weaker than others, the allowed
region of $X_\omega$ becomes wider.  Furthermore, we find that the CP
violating phases change importantly the prediction of $|\Theta_{\mu
  I}|$ even when the condition (\ref{eq:CR1_NH}) is satisfied.  
This is because, as found from Eq.~(\ref{eq:Amu_app}) in
App.~\ref{sec:appendix}, the negative $\xi \sin \eta$ decreases
$|\Theta_{\mu I}|$.  For instance, if we compare two sets of the CP
phases, ($\delta = \pi$, $\eta = \pi/2$) and ($\delta = 0$, $\eta = 3
\pi/2$), satisfying the condition (\ref{eq:CR1_NH}) with $\xi = +1$,
the latter case gives the suppressed $|\Theta_{\mu I}|$ for $X_\omega
\gg 1$ and $X_\omega$ can be about 350 as shown in
Fig.~\ref{fig:CONTOUR-NH}.%
\footnote{Here we have used the central values of parameters of
  active neutrinos.  When we vary them in the $3 \sigma$ range,
  $X_\omega$ can be large as about 550 as shown in Fig.~\ref{fig:AR}}

In the IH case, the mixing angle $\theta_{12}$ is an important
parameter to determine $\Theta_{\alpha I}$ and then we represent in
Fig.~\ref{fig:CONTOUR-IH} the upper bounds on $X_\omega$ in terms of
$\sin^2 \theta_{12}$.  Since the cancellation in $\Theta_{eI}$ cannot
be realized in the IH case, the experimental bounds on
$\abs{\Theta_{eI}}^2$ and $\abs{\Theta_{eI} \Theta_{\mu I}}$ place the
upper bound on $X_\omega$.  As explained in Eq.~(\ref{eq:Ratios_IH}),
the choice of ``$\xi \sin \eta$'' is significant to determine the
ratio $\abs{\Theta_{eI}}/\abs{\Theta_{\mu I}}$.  When $\xi \sin \eta =
-1$ (\eg, $\xi= + 1$ and $\eta = 3 \pi/2$), $\abs{\Theta_{eI}}$
becomes larger than $\abs{\Theta_{\mu I}}$.  In this case, the upper
bound on $X_\omega$ becomes smaller due to the stringent experimental
bound on $\abs{\Theta_{eI}}$ (see the left panel of
Fig.~\ref{fig:CONTOUR-IH}).  Inversely, when $\xi \sin \eta = +1$
(\eg, $\xi= + 1$ and $\eta = \pi/2$), $\abs{\Theta_{eI}}$ becomes
smaller than $\abs{\Theta_{\mu I}}$.  Then, the upper bound on
$X_\omega$ becomes relaxed and $X_\omega$ can be large as about 100
(see the right panel of Fig.~\ref{fig:CONTOUR-IH}).
 
We then find numerically the upper bound on $X_\omega$ for a given
$M_{2,3}$ by varying $\mbox{Re}\omega$, $\delta$ and $\eta$ in the
range $[0, 2\pi]$ and also by varying the parameters of active
neutrinos within the experimental 3$\sigma$ range.  The obtained
results are shown in Fig.~\ref{fig:AR}.  In the NH case, the upper
bound becomes weaker for lighter $M_N$.  Especially, no sensible 
bound is available for $M_N \lesssim 55$ MeV.  This is because the
cancellation in $\Theta_{eI}$ is possible and also because there is no
stringent experimental bound on $\abs{\Theta_{\mu I}}^2$ for such mass
region.  It is quite interesting to note that $X_\omega$ at ${\cal
  O}(10^2$-$10^3)$ is experimentally allowed for $M_N \sim 100$ MeV.
In the IH case, on the other hand, we can see that the bound on
$X_\omega$ is severer than that in the NH case.  This is because there
is no cancellation in $\Theta_{eI}$ and $X_\omega$ always receives the
stringent experimental bounds on $\abs{\Theta_{eI}}^2$ and/or
$\abs{\Theta_{eI} \Theta_{\mu I}}$ as explained above.

The upper bound of $X_\omega$ allows us to
estimate the possible range of the lifetimes of $N_{2}$ and $N_3$.
The lifetime is a key parameter from the following two reasons.  
One reason is concerned with the search for $N_2$ and $N_3$.  In the test of sterile
neutrino by finding charged leptons from its decay inside the
detector, the lifetime determines the decay length and 
also the detection rate.  The other comes from cosmology.
The light sterile neutrinos under consideration
can be long-lived particles
and their decays would spoil the success of the standard cosmology,
especially the prediction of light elements of BBN.
The lifetime is then restricted by various cosmological constraints.
We shall discuss this issue later.

In the considering mass region, $N_{2,3}$ decay in to $\nu_i \nu_j
\overline \nu_j$ or $\nu_i \ell^-_\alpha \ell^+_\beta$ ($\ell_{\alpha,
  \beta} = e, \mu$) and their CP conjugate states.  The lifetime,
which is determined from the mixing elements $\Theta_{\alpha I}$ and
the mass of sterile neutrino, can be estimated by using the partial
decay rates presented in
Refs.~\cite{Johnson:1997cj,Dolgov:2000jw,Gorbunov:2007ak}.  Before
dealing with the actual model, we consider the toy model described in
Eq.~(\ref{eq:TH_toy}).  In this case, since the dominant channel is
the decay into three active neutrinos, the lifetime is estimated as
\begin{eqnarray}
  \label{eq:life_toy}
  \tau_N \simeq
  \frac{192 \pi^3}{G_F^2 |\Theta|^2 M_N^5}
  =
  6 \times 10^3 \sec \left( \frac{m_{\rm atm}}{m_\nu} \right)
  \left( \frac{100 \MeV}{M_N} \right)^4 \,.
\end{eqnarray}
Therefore, sterile neutrino decays after BBN starts, which might lead
to cosmological difficulty.  In the $\nu$MSM, however, this is not
always the case and the much shorter lifetime can be achieved, because
the lifetime scales as $\tau_{N_{2,3}} \propto X_\omega^{-2}$ for
$X_\omega \gg 1$.  Thus, a sufficiently large $X_\omega$ makes
$N_{2,3}$ cosmologically harmless.

\begin{figure}[tb]
 \begin{minipage}{0.5\hsize}
   \centering
    \includegraphics[scale=1]{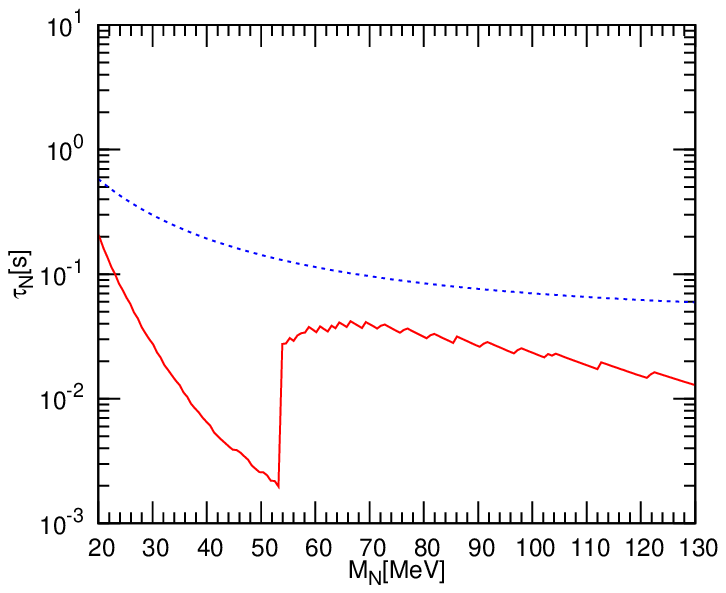}\\
 \end{minipage}
 \begin{minipage}{0.5\hsize}
   \centering
    \includegraphics[scale=1]{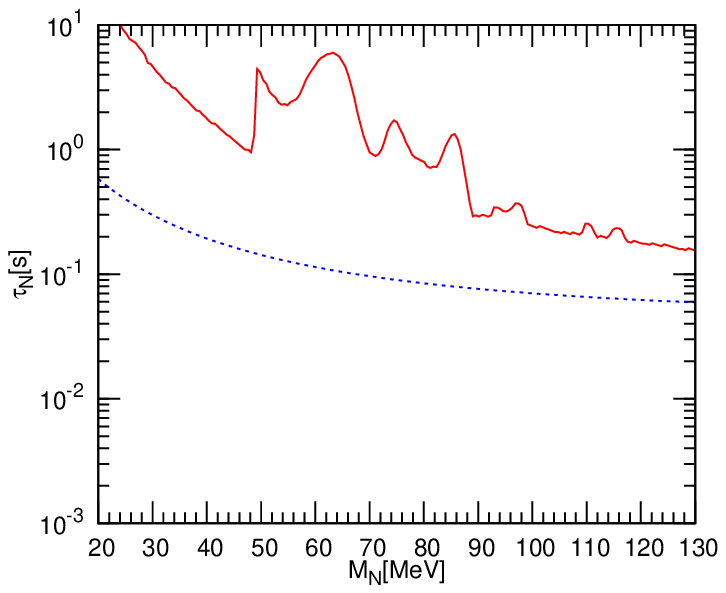}\\
 \end{minipage}
 \caption{Lower bounds on lifetime of sterile neutrino in terms of
   $M_N$ is shown by the red solid lines for the NH (left) and IH
   (right) cases.  We use the upper bound on the mixing elements for
   sterile neutrinos, $|\Theta_{e I}|^2_{\text{UB}}$, $|\Theta_{\mu
     I}|^2_{\text{UB}}$, $|\Theta_{e I} \Theta_{\mu I}|_{\text{UB}}$
   and the observational data for active neutrinos at the 3$\sigma$
   level.  The BBN upper bounds on lifetime in Eq.~(\ref{eq:tauBBN})
   are also shown by the blue dotted lines.
 }
 \label{fig:life}
\end{figure}
We find, as general properties, that the lifetimes of $N_2$ and $N_3$
are almost the same, and also that the lifetime in the IH case is
slightly shorter than the NH case for a given $X_\omega$.  
We then show the contour lines of
lifetime of $N_{2,3}$ in the parameter regions discussed previously
(see Figs.~\ref{fig:CONTOUR-NH}, \ref{fig:CONTOUR-IH} and
\ref{fig:AR}).  It is seen that the lifetime can vary orders of
magnitude depending on the choice of $X_\omega$.  Since we have
obtained the upper bound on $X_\omega$ to be consistent with the
search experiments, we can estimate the lowest value of the lifetime
of $N_{2,3}$.  The results are shown in Fig.~\ref{fig:life} for the NH
and IH cases, respectively.  In the NH case, the direct searches give
no significant constraint on $X_\omega$ for $M_N \lesssim 55$ MeV.  We
then impose $\abs{\Theta_{\alpha I}}^2 < 10^{-2}$ to ensure the
validity of the seesaw mechanism.  It is found that $\tau_{N_{2,3}}$
of ${\cal O}(10^{-2}$--$10^{-1})$ sec is available in the wide mass
range.  When the cancellation in $\Theta_{eI}$ occurs, the upper bound
on $X_\omega$ gets larger.  As a result, $|\Theta_{\mu I}|$ and
$|\Theta_{\tau I}|$ can take the larger values, which leads to the
shorter lifetime of $N_{2,3}$ even if $|\Theta_{eI}|$ is very small.
In the IH case, on the other hand, the absence of such a cancellation
gives the stringent lower bound of $\tau_{N_{2,3}}$ as 10--$10^{-1}$
sec for $M_N = 20$--130 MeV.

It had been pointed out in Ref.~\cite{Gorbunov:2007ak} that the bound
from the BBN places the upper bound on the lifetime of $N_{2,3}$,
which results in the admitted window for the parameter space of the
model.  The decays of sterile neutrinos around the BBN era would alter
the abundances of light elements in a desperate way.  There are two
significant effects by such decays.  One effect is that the 
additional energy carried by sterile neutrinos increases the expansion
rate of the universe.  The other is the modification of the
proton-neutron conversion rate.  This is because active neutrinos
produced by the decays of $N_{2,3}$ cannot be fully thermalized if the
decays occur too late and then the distortion of the distribution
functions affects the conversion rate.  These two effects would alter
the abundance of light elements ($^4$He especially) too much, which
leads to the upper bound on $\tau_{N_{2,3}}$.  Notice that
the stringent constraint from the BBN~\cite{Kawasaki:2004qu} 
on the hadronic decays can be avoided in the considering situation.

Such a bound had already been discussed in
Refs.~\cite{Dolgov:2000pj,Dolgov:2000jw}, in which the authors consider
the model with one flavour of active and sterile neutrinos.
At present there is no analysis dealing with the realistic situation
of the considering model.  
We shall postpone to study the issue in
future publication~\cite{AEI}, and just apply the bound on the
lifetime given in Ref.~\cite{Dolgov:2000jw}.  The maximum allowed
lifetime $\tau_{\rm BBN}$ is given by
\begin{eqnarray}
  \label{eq:tauBBN}
  \tau_{\rm BBN}/\mbox{sec} 
  = t_1 \, \left( \frac{M_N}{1\MeV} \right)^\beta + t_2\,,
\end{eqnarray}
where $t_1 = 128.7$, $t_2 = 0.04179$, and $\beta = -
1.828$~\cite{Dolgov:2000jw}.  (In this analysis we have taken a
conservative bound for the mixing with $\nu_{\mu, \tau}$, and 
see the detail in Ref.~\cite{Dolgov:2000jw}.)
When $M_N = 120$ MeV, we find $\tau_{\rm BBN} = 6.2 \times 10^{-2}$ sec.

\begin{figure}[t]
  \centerline{
  \includegraphics[scale=1.5]{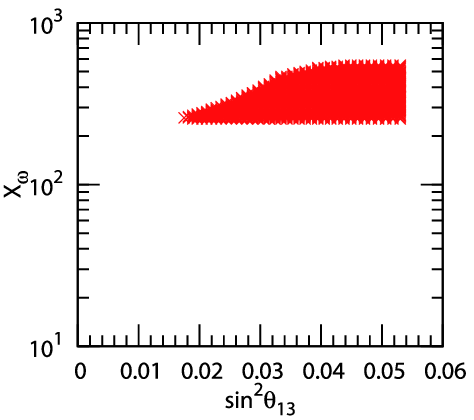}%
  \includegraphics[scale=1.5]{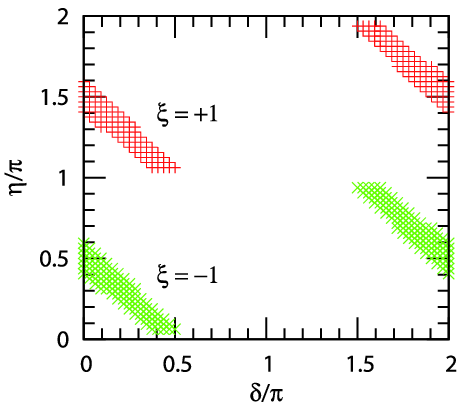}%
  }%
  \caption{Allowed regions in the $\sin^2 \theta_{13}$-$X_\omega$
    plane (left) and $\delta$-$\eta$ plane (right) when
    $\tau_{N_{2,3}} < \tau_{\rm BBN}$ for the NH case.  Here we take
    $M_N = 120$ MeV. In the right-panel red points and green points
    correspond to the case $\xi=+1$ and $\xi=-1$, respectively.}
  \label{fig:PARAM_NH_BBN}
\end{figure}
In Fig.~\ref{fig:life} the BBN bound on the lifetime
$\tau_{\rm BBN}$ is also shown in terms of $M_N$.  
We can find that, although such a short lifetime cannot be 
obtained in the IH case, there exits the allowed region 
in the NH case,  which is different from the conclusion in 
Ref.~\cite{Gorbunov:2007ak}.  This is because of the cancellation
in $\Theta_{eI}$ we have obtained in Sec.~\ref{sec:mixing},
which enlarges the allowed parameter space of the model
by relaxing the stringent constraints on $\abs{\Theta_{eI}}$
from direct searches.  In this allowed region, we can obtain 
remarkable predictions of the model parameters.
This point is illustrated in Fig.~\ref{fig:PARAM_NH_BBN}
when $M_N = 120$ MeV.
First, $X_\omega$ should be sufficiently large for 
$\tau_{N_{2,3}} < \tau_{\rm BBN}$ and $X_\omega \gtrsim 250$ in this case.
Second, in order to sufficiently suppress $\abs{\Theta_{eI}}$,
the parameters should be very close to those in the conditions
(\ref{eq:CR1_NH}) and (\ref{eq:CRt13_NH}).
Namely, the mixing angle $\theta_{13}$ should be large 
(close to $\theta_{13}^{\rm cr}$), and the CP violating phases
should be $\xi \sin (\delta + \eta) \simeq 1$.
Finally, the negative sign of $\xi \sin \eta$ is required
to decrease $\abs{\Theta_{\mu I}}$.
Therefore, the range of the Chooz angle $\theta_{13}$
and the CP phases $\delta$ and $\eta$ is highly restricted.
This leads to an important effect on baryogenesis 
of the model, which will be discussed elsewhere~\cite{AEI}.

In the following, however, we shall adopt the upper bound on the
lifetime as $\tau_{N_{2,3}} < 1$ sec, in addition to $\tau_{N_{2,3}} <
\tau_{\rm BBN}$, as the most conservative case, and consider the both
NH and IH cases.

Now, we are at the position to discuss the experimental search of
$N_{2,3}$ by the peak searches in the decays 
$\pi^+ \to e^+ + N_{2,3}$, $K^+ \to e^+ + N_{2,3}$,
and $K^+ \to \mu^+ + N_{2,3}$.
The branching ratio for the production process is given
by~\cite{Shrock:1980vy}
\begin{eqnarray}
  \mbox{BR}
  ( P^+ \to \ell^+ + N_I )
  =
  \mbox{BR}
  ( P^+ \to \ell^+ + \nu_\ell )
  \times K( m_{P}, m_\ell, M_I ) \,,
\end{eqnarray}
where $P = \pi, K$ and $\ell = e, \mu$. The function
$K$ is 
\begin{eqnarray}
  K( m_P, m_\ell, M_I)
  = 
  \abs{\Theta_{\ell I}}^2 \,
  \frac{ \beta_{N_I} }{ \beta_\nu } \,
  \frac{ m_P^2 ( M_I^2 + m_\ell^2 ) - (M_I^2 - m_\ell^2)^2}
  { m_P^2 m_\ell^2 - m_\ell^4 } \,,
\end{eqnarray}
where 
\begin{eqnarray}
  \beta_{N_I}^2 = 
  1 - 2 \frac{M_I^2 + m_\ell^2}{m_P^2}  
  + \frac{ ( M_I^2 - m_\ell^2 )^2 }{ m_P^4 } \,,
  ~~~~~
  \beta_\nu =
  1 - \frac{ m_\ell^2 }{ m_P^2 } \,.
\end{eqnarray}
Here we have neglected the masses of active neutrinos.
It is then seen that the branching ratios are proportional
to the mixing element squared, and hence $X_\omega^2$.

\begin{figure}[t]
  \centerline{
  \includegraphics[scale=1.1]{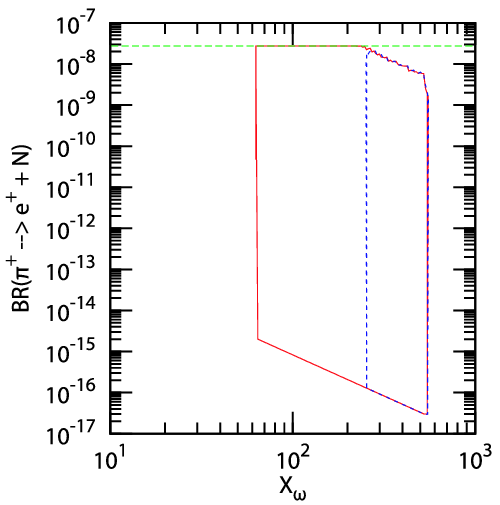}%
  \includegraphics[scale=1.1]{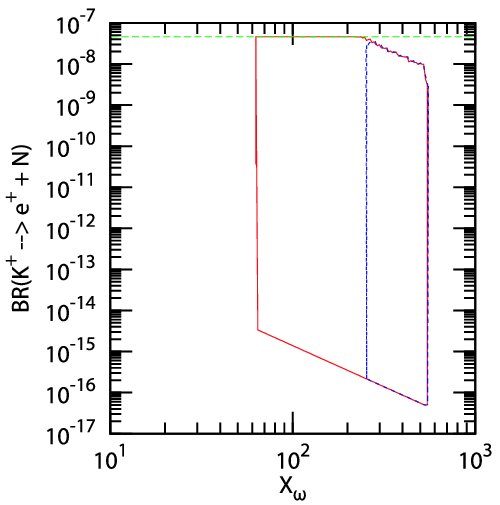}%
  \includegraphics[scale=1.1]{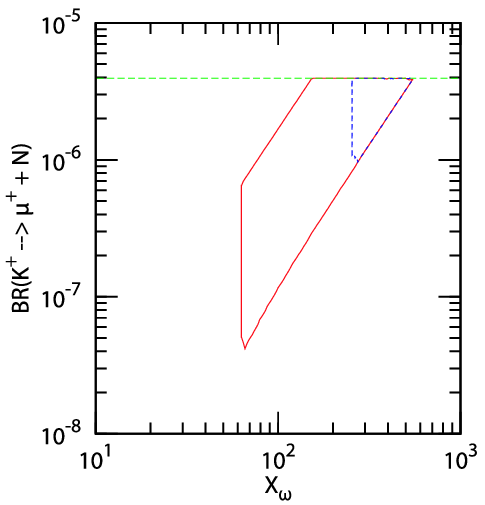}%
  }%
  \caption{Allowed regions for the branching ratios of $\pi^+ \to e^+
    + N_{2,3}$ (left), $K^+ \to e^+ + N_{2,3}$ (center), and $K^+ \to
    \mu^+ + N_{2,3}$ (right) in terms of $X_\omega$ for the NH case.
    The regions within the red solid lines and the blue dotted lines are
    allowed for the case when $\tau_{N_{2,3}} < 1$ sec and
    $\tau_{N_{2,3}} < \tau_{\rm BBN}$, respectively.  The green
    horizontal lines represent the branching ratios when $|\Theta_{eI}|
    = |\Theta_{eI}|_{\rm UB}$ (left and center) and $|\Theta_{\mu I}|
    = |\Theta_{\mu I}|_{\rm UB}$ (right).  }
  \label{fig:BRXom_NH}
\end{figure}
%
\begin{figure}[ht]
  \centerline{
  \includegraphics[scale=1.1]{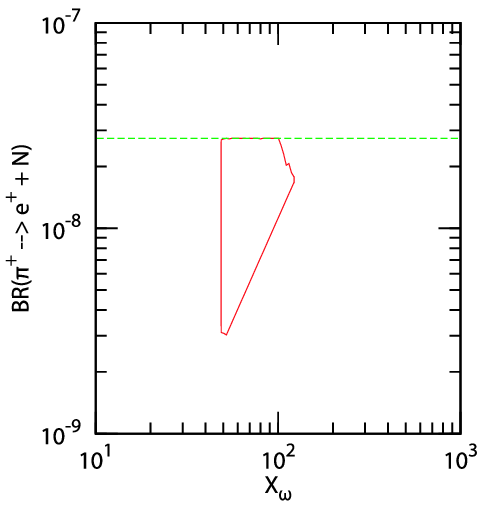}%
  \includegraphics[scale=1.1]{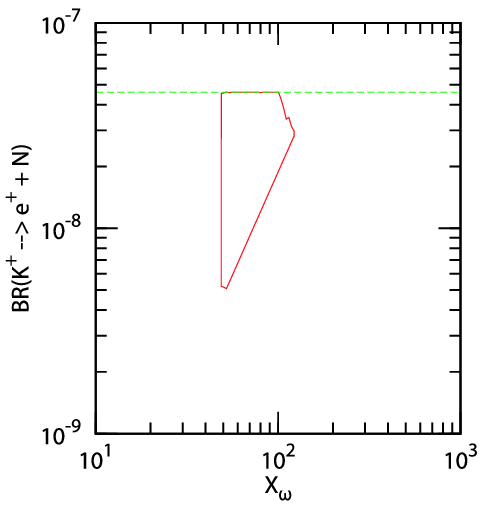}%
  \includegraphics[scale=1.1]{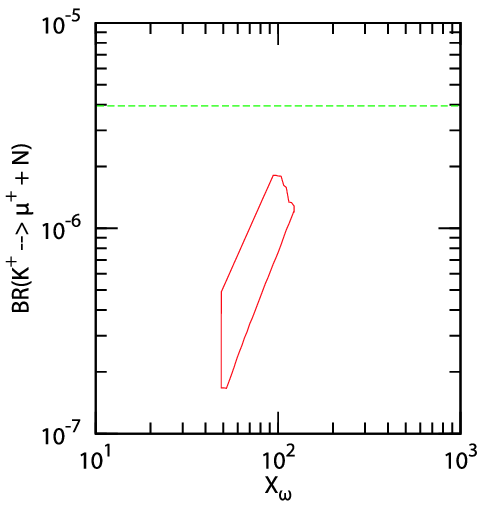}%
  }%
  \caption{Allowed regions for the branching ratios of $\pi^+ \to e^+
    + N_{2,3}$ (left), $K^+ \to e^+ + N_{2,3}$ (center), and $K^+ \to
    \mu^+ + N_{2,3}$ (right) in terms of $X_\omega$ for the IH case.
    The regions with the red solid lines are allowed for the case when
    $\tau_{N_{2,3}} < 1$ sec.  The green horizontal lines represent
    the branching ratios when $|\Theta_{eI}| = |\Theta_{eI}|_{\rm UB}$
    (left and center) and $|\Theta_{\mu I}| = |\Theta_{\mu I}|_{\rm
      UB}$ (right).}
  \label{fig:BRXom_IH}
\end{figure}
By taking $M_N = 120$ MeV, the allowed regions of the branching ratios
are shown in Figs.~\ref{fig:BRXom_NH} and
\ref{fig:BRXom_IH} for the NH and IH cases, respectively.  
Here we have imposed the lifetime bound in addition to 
the experimental bounds on the mixing elements,
and the parameters of the Yukawa matrix have been 
varied as we did before.

In the NH case, it is found that the branching ratios of $\pi^+ \to
e^+ + N$ and $K^+ \to e^+ + N$ can be very small even if one imposes
the lifetime bound.  This is due to the cancellation in
$|\Theta_{eI}|$ pointed out in this paper.  As explained in
Eq.~(\ref{eq:THesup_NH}), $|\Theta_{eI}|^2$ is proportional to
$X_\omega^{-2}$ when the conditions (\ref{eq:CR1_NH}) and
(\ref{eq:CRt13_NH}) are satisfied, and hence the lower bounds on these
branching ratios are also proportional to $X_\omega^{-2}$.  
For large values
of $X_\omega$ the upper bounds on these branching ratios are smaller
than those evaluated by using the experimental upper bound
$|\Theta_{eI}|^2_{\rm UB}$.  This is because $|\Theta_{\mu I}|^2$ must
take its maximal value $|\Theta_{\mu I}|^2_{\rm UB}$ to realize such a
large $X_\omega$, and then $|\Theta_{e I} \Theta_{\mu I}|_{\rm UB}$ in
addition to $|\Theta_{\mu I}|^2_{\rm UB}$ forbids the case
$|\Theta_{eI}|^2 = |\Theta_{eI}|^2_{\rm UB}$.  On the other hand, the
branching ratio of $K^+ \to \mu^+ + N_{2,3}$ cannot be so small since
there is no cancellation in $|\Theta_{\mu I}|$
and also the larger value of $|\Theta_{\mu I}|$ is required for
the shorter lifetime.  Notice that
$|\Theta_{\mu I}|^2$ cannot be large as
$|\Theta_{\mu I}|^2_{\rm UB}$ 
for smaller values of $X_\omega$.
This is the reason why the upper bound on 
$\mbox{BR}(K^+ \to \mu^+ + N_{2,3})$ 
is smaller than that evaluated by 
$|\Theta_{\mu I}|^2 = |\Theta_{\mu I}|^2_{\rm UB}$ when $X_\omega \lesssim 150$. 

Therefore, we arrive at significant conclusions
on the direct searches of $N_{2,3}$ in the NH case.  By the peak
search of positrons in $\pi^+$ or $K^+$ decays we would miss $N_2$ and
$N_3$, since the model predicts too small branching ratios of these
decays in some cases.  Further, even if this is the case, it is
possible to detect $N_{2,3}$ in the $\mu^+$ spectrum in the $K^+$
decays.  Interestingly, the future experiments of $\mbox{BR}(K^+ \to
\mu^+ + N_{2,3})$ improved the sensitivity by two orders of magnitude
can cover the whole parameter space when $\tau_{N_{2,3}} < 1$ sec.

In the IH case, as shown in Fig.~\ref{fig:BRXom_IH}, the regions of
the branching ratios are different from those in the NH case.
Especially, the branching ratios, $\pi^+ \to e^+ + N_{2,3}$ and $K^+
\to e^+ + N_{2,3}$ cannot be so small as in the NH case 
due to the absence of the
cancellation in $|\Theta_{eI}|$.  Further, it should
be noted that the model predict the branching ratio of $K^+ \to \mu^+
+ N_{2,3}$ below the value with $|\Theta_{\mu I}|^2 = |\Theta_{\mu
  I}|^2_{\rm UB}$ for all the range of $X_\omega$.  The reason for
small values of $X_\omega$ is the same as the NH case.  In addition,
the ratio $|\Theta_{eI}|/|\Theta_{\mu I}|$ cannot be very small in the
IH case, and then the stringent experimental bound
$|\Theta_{eI}|^2_{\rm UB}$ forbids $|\Theta_{\mu I}|^2$ being large as
$|\Theta_{\mu I}|^2_{\rm UB}$.  This leads to the experimental
signature of the IH case, \ie, the suppressed branching
ratio of $K^+ \to \mu^+ + N_{2,3}$.  In the IH case, therefore, the
improvements of the experiments by one order of magnitude
can cover the whole parameter range when $\tau_{N_{2,3}} < 1$ sec for
all the three decay channels.

\begin{figure}[!t]
  \centerline{
  \includegraphics[scale=1.1]{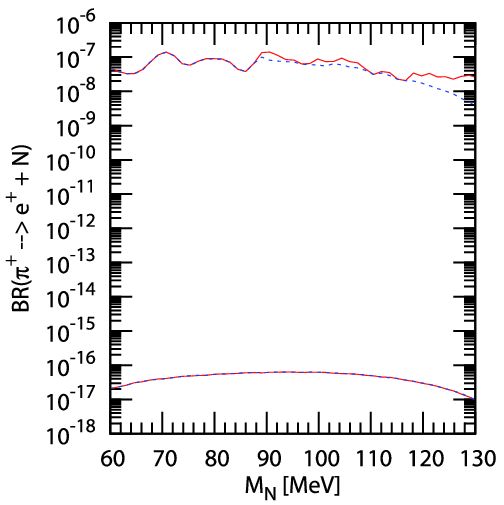}%
  \includegraphics[scale=1.1]{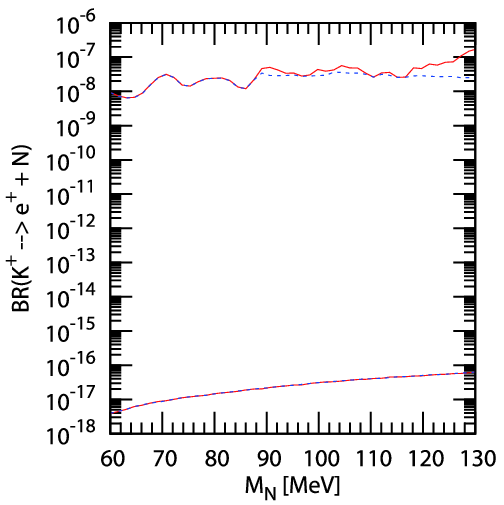}%
  \includegraphics[scale=1.1]{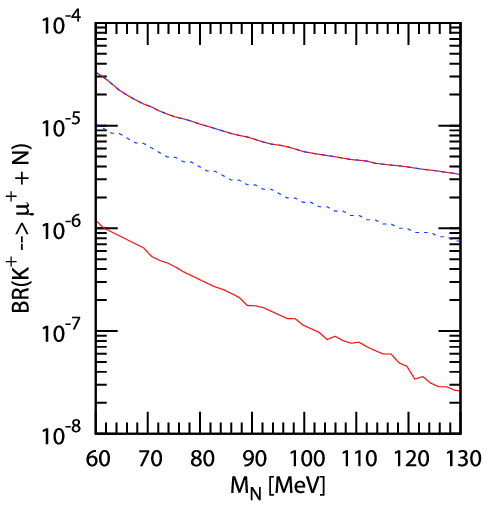}%
  }%
  \caption{Allowed regions of
    the branching ratios of $\pi^+ \to e^+ + N_{2,3}$ (left),
    $K^+ \to e^+ + N_{2,3}$ (center), and $K^+ \to \mu^+ + N_{2,3}$
    (right) in terms of $M_N$ for the NH case.
    The regions within the red solid lines 
    and the blue dotted lines are allowed
    for the case when $\tau_N < 1$ sec and 
    $\tau_N < \tau_{\rm BBN}$~(\ref{eq:tauBBN}).
  }
  \label{fig:BRMN_NH}
  \vspace{5ex}
  \centerline{
  \includegraphics[scale=1.1]{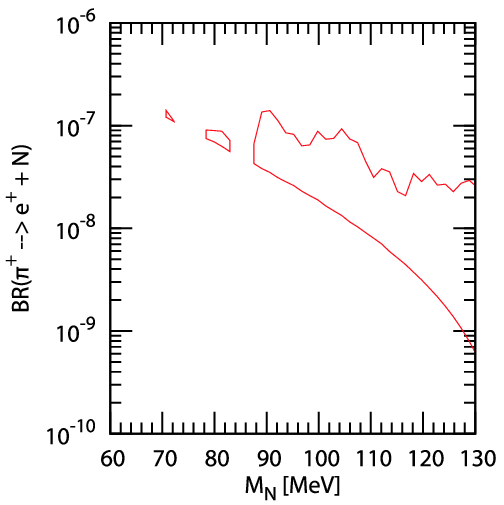}%
  \includegraphics[scale=1.1]{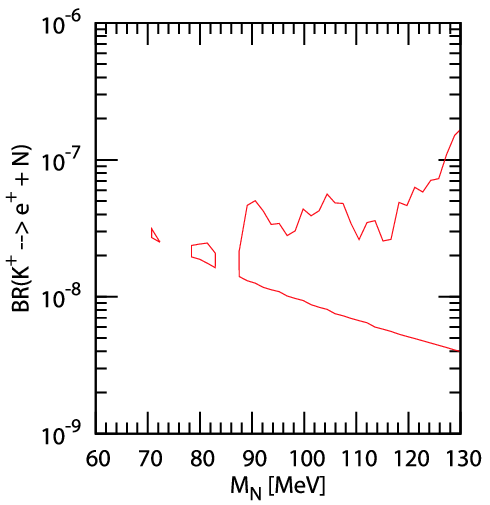}%
  \includegraphics[scale=1.1]{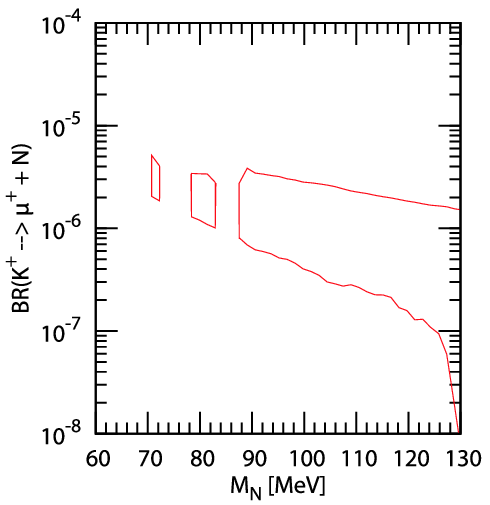}%
  }%
  \caption{Allowed regions of 
    the branching ratios of $\pi^+ \to e^+ + N_{2,3}$ (left),
    $K^+ \to e^+ + N_{2,3}$ (center), and $K^+ \to \mu^+ + N_{2,3}$
    (right) in terms of $M_N$ for the IH case.  
      The regions within the red solid lines are allowed
    with $\tau_N < 1$ sec.}
  \label{fig:BRMN_IH}
\end{figure}
Finally, we show in Figs.~\ref{fig:BRMN_NH} and \ref{fig:BRMN_IH} the
allowed ranges of the branching rations in terms of $M_N$ for the NH
and IH cases, respectively.  Here we also impose the lifetime bound
and vary the parameters of the model including $X_\omega$.  In the NH
case, we find that the lower bounds on the branching ratios $\pi^+ \to
e^+ + N_{2,3}$ and $K^+ \to e^+ + N_{2,3}$ are the same between two
cases when $\tau_{N_{2,3}} < 1$ sec and $\tau_{N_{2,3}} < \tau_{\rm
  BBN}$.  This is because such bounds are obtained by the largest
$X_\omega$ as shown in Fig.~\ref{fig:BRXom_NH}.  It is also find that
the upper bound for $\tau_{N_{2,3}} < \tau_{\rm BBN}$ is slightly
smaller than that for $\tau_{N_{2,3}} < 1$ sec.  On the other hand,
the upper bound on the branching ratio of $K^+ \to \mu^+ + N_{2,3}$
does not change by the lifetime bound.  In contrast, the lower bound
becomes much severer by the stringent lifetime bound $\tau_{N_{2,3}} <
\tau_{\rm BBN}$.  Therefore, the improvement in
the measurement of $\mbox{BR}(K^+ \to
\mu^+ + N_{2,3})$ by a factor of 5 (by ${\cal O}(10^2)$)
is sufficient to cover the
allowed region with $\tau_{N_{2,3}} < \tau_{\rm BBN}$ 
($\tau_{N_{2,3}} < 1$ sec) for 
$60\MeV \lesssim M_N \lesssim 130$ MeV.
In the IH case, for all three decay channels, the improvement in the
measurements by one or two orders of magnitude allows to cover the
allowed region for $60\MeV \lesssim M_N \lesssim 130$ MeV.

Before closing this section, we would like to remark the following two
points.  First, the properties of the mixing elements $\Theta_{\alpha
  I}$ obtained in Sec.~\ref{sec:mixing} are independent of the choice
of the masses $M_{2,3}$.  Therefore, the above arguments can be
applied to sterile neutrinos heavier than pion mass.  
It should be stressed in particular that
we have to pay a special attention to the processes of
$N_{2,3}$ associated with electron/positron because of the
cancellation in $\Theta_{eI}$ for the NH case.  Second, we would like
to comment again that, although we consider the $\nu$MSM in this
paper, most of the results still hold for the general models of the
seesaw mechanism with two right-handed neutrinos.
\section{Conclusions}
\label{sec:conc}
We have discussed mixing of active and sterile neutrinos in the
$\nu$MSM, paying special attention to the mixing elements,
$\Theta_{\alpha 2}$ and $\Theta_{\alpha 3}$, of $N_2$ and $N_3$.  In
this model, these sterile neutrinos are responsible to generate the
seesaw mass matrix of active neutrinos as well as the baryon asymmetry
of the universe through the mechanism of neutrino oscillation.  Since
these mixing elements are crucial to determine the strength of
interaction of $N_{2,3}$, we have investigated the properties of
$\Theta_{\alpha 2}$ and $\Theta_{\alpha 3}$ in detail.

It has been shown that the parameter $\mbox{Im}\omega$ (or $X_\omega$)
is important to determine the overall scale of the Yukawa couplings of
neutrinos.  The couplings of $N_{2,3}$ scales as $|F_{\alpha I}|
\propto e^{|{\rm Im}\omega |}$ and changes by orders of magnitude for
$|\mbox{Im} \omega| \gg 1$.  We should stress again that the choice of
$\mbox{Im} \omega$ does not change the masses and mixing angles of
active neutrinos.  Thus, the mixing elements $|\Theta_{\alpha I}|$ of
$N_{2,3}$ can be larger as $|\mbox{Im} \omega|$ becomes large, being
consistent with the oscillation data.

We have then presented how the mixing elements depend on the neutrino
parameters, namely masses, mixing angles and CP phases of active
neutrinos.  Interestingly, we have observed in the NH case that the
leading terms of $|\Theta_{e2}|$ and $|\Theta_{e3}|$ vanish at the
same time and they becomes suppressed as $X_\omega^{-2}$ for the large
$X_\omega$ region, if the conditions (\ref{eq:CR1_NH}) and
(\ref{eq:CRt13_NH}) are satisfied.  This is our important result,
which is illustrated in Fig.~\ref{fig:FIG_THSQ_NH}.  In order to
realize this cancellation or the strong suppression in $|\Theta_{e2}|$
and $|\Theta_{e3}|$, the large value of $\theta_{13}$ is required.
Thus, the experiments, \eg, Double Chooz, T2K, RENO, Daya Bay, and
NO$\nu$A in near future~\cite{Mezzetto:2010zi} will check whether it
can happen in nature or not.  The similar cancellation or suppression
in $|\Theta_{e2}|$ and $|\Theta_{e3}|$ are potentially possible in the
IH case.  However, we have shown that the required conditions cannot
be satisfied within the current neutrino data with $3 \sigma$ error.

We have also discussed the $0 \nu 2 \beta$ decays in the $\nu$MSM.
Both contributions from active neutrinos and sterile neutrinos are
fully taken into account.  It has been found that quasi-degenerate
sterile neutrinos $N_2$ and $N_3$ generating the baryon asymmetry of
the universe gives a significant effect when their masses are smaller
than about 100 MeV which is the typical energy scale in the nuclear
matrix elements of the decays.  In fact, when their masses are well
below 100 MeV, the effective neutrino mass in the $0 \nu 2 \beta$
decays vanishes in the $\nu$MSM, as already pointed out in
Ref.~\cite{Blennow:2010th} by using the general seesaw model.
Moreover, when their masses are comparable to 100MeV, sterile
neutrinos, $N_2$ and $N_3$, give destructive contribution and then
$m_{\rm eff}$ becomes much smaller than that from active neutrinos.
In this case, therefore, the $0 \nu 2 \beta$ decays are suppressed due
to the existence of light sterile neutrinos.  Inversely speaking, the
constraints from the $0 \nu 2 \beta$ decays on the parameters of the
$\nu$MSM become weaker in this mass region.  On the other hand, when
$N_2$ and $N_3$ are much heavier than 100 MeV, the contributions to
the $0 \nu 2 \beta$ decays from sterile neutrinos are negligible and
$m_{\rm eff}$ is mainly comes from three active
neutrinos as in the conventional seesaw model.

Moreover, we have estimated the lifetimes of $N_2$ and $N_3$ with
$M_{2,3} < m_{\pi}$.  It has been found that the lifetime can be
shorter by taking large $X_\omega$ due to the enhancements of the
mixing elements.  As shown in Fig.~\ref{fig:life}, the present
experimental limits on the mixing elements pose the lower bound on the
lifetime as $10^{-2}$ sec at $M_N \simeq 120$ MeV in the NH case and
$10^{-1}$ sec at $M_N \simeq 120$ MeV in the IH case.  The obtained
result is essential for the discussion of the cosmological constraints
on the decays of $N_2$ and $N_3$.  Especially, we have pointed out
that, even if one imposes the current cosmological bound from the BBN
as $\tau_{N_{2,3}}< \tau_{\rm BBN}$, there exists an allowed region
for such light sterile neutrinos in the NH case.  This is because of
the cancellation in $\abs{\Theta_{eI}}$ obtained in the present
analysis.  
We have then derived the predictions on neutrino parameters,
namely, $\theta_{13}$ is large,
$\xi \sin (\delta + \eta ) \simeq -1$, and
$\xi \sin \eta < 0$.
When we take the weaker bound as $\tau_{N_{2,3}} < 1$ sec,
there are the allowed regions for both NH and IH cases.

Finally, we have investigated the direct search of $N_2$ and $N_3$
with $M_{2,3} < m_{\pi}$ in meson decays.  Our study shows that, in
the NH case, the experiments by using $\pi^+ \to e^+ + N_{2,3}$ 
and $K^+ \to e^+ + N_{2,3}$ 
would miss $N_{2,3}$ when $\theta_{13}$ is large.  This is
because the cancellation or strong suppression in $\Theta_{eI}$ can
happen.  In such a case, the peak search of the kaon decays $K^+ \to
\mu^+ + N_{2,3}$ is crucially important to find $N_{2,3}$ having
sizable mixing elements $\Theta_{\mu I}$.  We should also comment that
$N_{2,3}$ with $M_N \lesssim 55$ MeV is hard to find in the NH case,
if the cancellation in $\Theta_{eI}$ happens.  This is because the
peak search in $K^+ \to \mu^+ + N_{2,3}$ is difficult for such a 
small mass region.  On the other hand, in the IH case, 
all the three decay channels,
$\pi^+ \to e^+ + N_{2,3}$, $K^+ \to e^+ + N_{2,3}$, and
$K^+ \to \mu^+ + N_{2,3}$, are promising 
for the future experiments finding $N_2$ and $N_3$.

\section*{Acknowledgments}
We would like to thank M.~Shaposhnikov for useful discussions.
The work of T.A. was partially supported by the Ministry of Education,
Science, Sports and Culture, Grant-in-Aid for Scientific Research,
No.~21540260, and by Niigata University Grant for Proportion of
Project.

\appendix
\section{Mixing elements $|\Theta_{\alpha I}|^2$ at ${\cal O}(X_\omega^2)$}
\label{sec:appendix}
In this appendix, we shall present the expressions of the mixing
elements $|\Theta_{\alpha I}|^2$ at the ${\cal O}(X_\omega^2)$.  In
the NH case, we parameterize $|\Theta_{\alpha I}|^2$ at the ${\cal
  O}(X_\omega^2)$  as
\begin{eqnarray}
  \left. 
    |\Theta_{\alpha I}|^2 
  \right|_{X_\omega^2}
  =  \frac{X_\omega^2 \, m_3}{4 M_I} \, A_\alpha \,,
\end{eqnarray}
where $A_\alpha$ ($\alpha = e, \mu, \tau$) are found as
[$r_{23} = (m_2/m_3)^{1/2}$]
\begin{eqnarray}
  A_e 
  \eqn{=}
  \cos^2 \theta_{13} 
  \Bigl[ \tan^2 \theta_{13} 
    + 2 r_{23} \xi \sin \left(\delta + \eta \right)  
    \sin \theta_{12} \tan \theta_{13} 
    + r_{23}^2 \, \sin^2 \theta_{12} 
  \Bigr] \,,  
  \\
  \label{eq:Amu_app}
  A_\mu 
  \eqn{=} 
  \cos^2 \theta_{13} \sin^2 \theta_{23} 
  +
  r_{23} \, \xi 
  \Bigl[
  \cos \theta_{12} \cos \theta_{13} \sin 2 \theta_{23} \sin \eta
  -
  \sin \theta_{12} \sin 2\theta_{13} \sin^2  \theta_{23} \sin (\delta + \eta)
  \Bigr]
  \nonumber \\
  \eqn{} 
  +
  r_{23}^2
  \Bigl[
  \cos^2 \theta_{12} \cos^2 \theta_{23}
  + 
  \sin^2 \theta_{12} \sin^2 \theta_{13} \sin^2 \theta_{23}
  - \frac{1}{2} \sin 2 \theta_{12} \sin 2 \theta_{23} \sin \theta_{13}
  \cos \delta 
  \Bigr] \,,
  \\
  A_\tau
  \eqn{=} 
  \cos^2 \theta_{13} \cos^2 \theta_{23} 
  -
  r_{23} \, \xi 
  \Bigl[
  \cos \theta_{12} \cos \theta_{13} \sin 2 \theta_{23} \sin \eta
  +
  \sin \theta_{12} \sin 2\theta_{13} \cos^2  \theta_{23} \sin (\delta + \eta)
  \Bigr]
  \nonumber \\
  \eqn{} 
  +
  r_{23}^2
  \Bigl[
  \cos^2 \theta_{12} \sin^2 \theta_{23}
  + 
  \sin^2 \theta_{12} \sin^2 \theta_{13} \cos^2 \theta_{23}
  + \frac{1}{2} \sin 2 \theta_{12} \sin 2 \theta_{23} \sin \theta_{13}
  \cos \delta 
  \Bigr] \,.
\end{eqnarray}
On the other hand, in the IH case, we 
\begin{eqnarray}
  \left. 
    |\Theta_{\alpha I}|^2 
  \right|_{X_\omega^2}
  =  \frac{X_\omega^2 \, m_2}{4 M_I} \, B_\alpha \,,
\end{eqnarray}
where $B_\alpha$ ($\alpha = e, \mu, \tau$) are found as
[$r_{12} = (m_1/m_2)^{1/2}$]
\begin{eqnarray}
  B_e 
  \eqn{=} 
  \cos^2 \theta_{12} \cos^2 \theta_{13}
  \Bigl[
  \tan^2 \theta_{12} - 2 r_{12} \xi \sin \eta \tan \theta_{12} + r_{12}^2
  \Bigr]\,,
\end{eqnarray}
\begin{eqnarray}
  B_\mu
  \eqn{=}
  \cos \theta_{12}^2 \cos^2 \theta_{23}
  + \sin^2 \theta_{12} \sin \theta_{13}^2 \sin^2 \theta_{23}
  - \frac{1}{2} 
  \sin 2 \theta_{12} \sin 2 \theta_{23} \sin \theta_{13} \cos \delta
  \nonumber \\
  \eqn{}
  + r_{12} \xi
  \Bigl[
  ( \cos 2 \theta_{12} \cos \delta \sin \eta - \sin \delta \cos \eta )
  \sin 2 \theta_{23} \sin \theta_{13}
  \nonumber \\
  \eqn{}~~~~~~~~~
  +
  \sin 2 \theta_{12} 
  ( \cos^2 \theta_{13} \sin^2 \theta_{23} + \cos 2 \theta_{23} )
  \sin \eta
  \Bigr]
  \nonumber \\
  \eqn{}
  + r_{12}^2 
  \Bigl[
  \cos^2 \theta_{23} \sin^2 \theta_{12}
  +
  \cos^2 \theta_{12} \sin^2 \theta_{13} \sin^2 \theta_{23}
  +
  \frac{1}{2} \sin 2 \theta_{12} \sin 2 \theta_{23} \sin \theta_{13}
  \Bigr] \,,
\end{eqnarray}
\begin{eqnarray}
  B_\tau
  \eqn{=}
  \cos \theta_{12}^2 \sin^2 \theta_{23}
  + \sin^2 \theta_{12} \sin \theta_{13}^2 \cos^2 \theta_{23}
  + \frac{1}{2} 
  \sin 2 \theta_{12} \sin 2 \theta_{23} \sin \theta_{13} \cos \delta
  \nonumber \\
  \eqn{}
  + r_{12} \xi
  \Bigl[
  (- \cos 2 \theta_{12} \cos \delta \sin \eta + \sin \delta \cos \eta )
  \sin 2 \theta_{23} \sin \theta_{13}
  \nonumber \\
  \eqn{}~~~~~~~~~
  +
  \sin 2 \theta_{12} 
  ( \cos^2 \theta_{13} \cos^2 \theta_{23} - \cos 2 \theta_{23} )
  \sin \eta
  \Bigr]
  \nonumber \\
  \eqn{}
  + r_{12}^2 
  \Bigl[
  \sin^2 \theta_{23} \sin^2 \theta_{12}
  +
  \cos^2 \theta_{12} \sin^2 \theta_{13} \cos^2 \theta_{23}
  -
  \frac{1}{2} \sin 2 \theta_{12} \sin 2 \theta_{23} \sin \theta_{13}
  \Bigr] \,.
\end{eqnarray}

\end{document}